\documentclass[nohyper,12pt,letterpaper]{JHEP3}
\usepackage{graphicx,epsfig,youngtab}

\newcommand{\myfig}[3]{\begin{figure}[ht]
\begin{center}
\leavevmode \epsfxsize=#2cm \epsfbox{#1}
\end{center}
\caption{#3} \label{fig:#1}
\end{figure}}

\author{Rajsekhar Bhattacharyya$^{1}$, Robert de Mello Koch$^{2,3}$ and Michael Stephanou$^{2}$\\
\qquad \\
$^{1}$ Department of Physics,\\
Dinabandhu Andrews College,\\
Kolkata-700084,India\\
$^{2}$ National Institute for Theoretical Physics,\\
Department of Physics and Centre for Theoretical Physics,\\ 
University of the Witwatersrand,\\ 
Wits, 2050, South Africa\\
$^{3}$Stellenbosch Institute for Advanced Studies,\\
Stellenbosch, South Africa\\
\qquad\\
E-mail: \email{rajsekhar@dacollege.org, robert@neo.phys.wits.ac.za, michael.stephanou@students.wits.ac.za}}

\abstract{We derive a product rule satisfied by restricted Schur polynomials. We focus mostly on the case that the
restricted Schur polynomial is built using two matrices, although our analysis easily extends to more than two
matrices. This product rule allows us to
compute exact multi-point correlation functions of restricted Schur polynomials, in the free field theory limit.
As an example of the use of our formulas, we compute two point functions of certain single trace operators built
using two matrices and three point functions of certain restricted Schur polynomials, exactly, in the free field
theory limit. Our results suggest that gravitons become strongly coupled at sufficiently high energy, while the
restricted Schur polynomials for totally antisymmetric representations remain weakly interacting at these energies. 
This is in perfect accord with the half-BPS (single matrix) results of hep-th/0512312. Finally, by studying the 
interaction of two restricted Schur polynomials we suggest a physical interpretation for the labels of the restricted
Schur polynomial: the composite operator $\chi_{R,(r_n,r_m)}(Z,X)$ 
is constructed from the half BPS ``partons'' $\chi_{r_n}(Z)$ and $\chi_{r_m}(X)$.}

\preprint{WITS-CTP-037}

\title{Exact Multi-Restricted Schur Polynomial Correlators}


\def \Tr{\mbox{Tr\,}}

\begin{document}

\section{Introduction}

The AdS/CFT correspondence\cite{Maldacena:1997re} has provided the possibility of studying quantum gravity non-perturbatively.
A good example of this progress are the half-BPS type IIB geometries in asymptotically AdS spaces, constructed by 
Lin, Lunin and Maldacena\cite{Lin:2004nb} and their connection with the semiclassical states of a free fermi 
system\cite{Corley:2001zk,Berenstein:2004kk}. 
It would be fascinating to extend these studies beyond the half-BPS sector. The present work is directed at this goal.

In the boundary super Yang-Mills theory, the operators relevant for the half-BPS LLM geometries are constructed
using a single free complex Higgs field $Z$. The dynamics of a single free matrix is captured by $N$ free fermions
in a harmonic potential\cite{Corley:2001zk,Berenstein:2004kk,Brezin:1977sv,related,Takayama:2005yq}. At large $N$ 
this single matrix dynamics has a semiclassical description in terms of droplets of fermi fluid in phase space. 
Remarkably, LLM showed that there is a similar structure in the classical geometries in the bulk.

To go beyond the half-BPS sector, one needs to study multi-matrix dynamics. In general, this is a formidable problem.
There has however, been some recent progress: three independent bases for general multi-matrix models have been identified.
For a review of these developments and the work leading up to them, see \cite{sanjaye}.
The  basis described in \cite{Brown:2007xh} builds operators with definite flavor quantum numbers; we call this basis the
flavor basis. The basis of \cite{Kimura:2007wy} uses the Brauer algebra to build correlators involving $Z$ and $Z^\dagger$;
we call this basis the Brauer basis. The Brauer basis seems to be the most natural for exploring brane/anti-brane systems.
The basis of \cite{Bhattacharyya:2008rb} most directly allows one to consider 
open string excitations\cite{Balasubramanian:2002sa,Balasubramanian:2004nb,de Mello Koch:2007uu,de Mello Koch:2007uv,Bekker:2007ea} 
of the operator; we call this the restricted Schur basis. These three bases do not coincide and a detailed link between them 
is an interesting open problem. All three bases diagonalize the two point functions in the free field theory limit.

Completeness of the flavor basis was convincingly demonstrated\cite{Brown:2007xh} by arguing that the number of such operators 
matches the number of gauge invariant operators that can be constructed\cite{Dolan:2007rq}, at both infinite and finite $N$. The 
flavor basis also gives a group theoretic way to approach higher point functions (see \cite{Brown:2007xh} where three and higher point 
functions are obtained) and to obtain factorization equations which can be used to build a probability interpretation\cite{Brown:2006zk}. 
By exploiting supergroups \cite{Brown:2007xh} have also explained how to include fermions in addition to the Higgs fields.
Finally, the one loop correction to these two points functions has been considered in \cite{Brown:2008rs}.

Arguments for the completeness of the restricted Schur basis in \cite{Bhattacharyya:2008rb} demonstrated the number of restricted
Schur polynomials matches the number of gauge invariant operators that can be constructed\cite{Dolan:2007rq}, again at both infinite 
and finite $N$, for a number of examples. A proof of this matching for infinite $N$ is known\cite{Brown}. In this article we will 
compute exact three and higher point correlation functions of restricted Schur polynomials, in the free field theory limit.

Our approach to the computation of multipoint functions has a simple algebraic description: We start by deriving a product rule
on the space of restricted Schur polynomials: the product of any two restricted Schur polynomials can be expressed as a linear
combination of restricted Schur polynomials. Applying the rule $(n-1)$ times we can collapse the product of $n$ restricted
Schur polynomials to a linear combination of restricted Schur polynomials. In this way, an arbitrary multipoint function can
be reduced to a linear combination of two point functions - something that we know how to compute. These multipoint functions
contain dynamical information about the theory, and hence they should reveal interesting physics. 
As an example,
studying the Yang-Mills theory at finite $N$ probes truly quantum mechanical aspects of the bulk gravity.
At finite $N$ only fluctuations with a low enough excitation energy can be interpreted as gravity modes propagating
in the bulk. It is appropriate, at low energies, to use the graviton degrees of freedom to set up a perturbative description. At high
energies, the description will employ more fundamental microscopic degrees of freedom. Since
$$ {R\over l_p}\sim N^{1\over 4},$$
with $R$ the AdS radius and $l_p$ the ten-dimensional Planck scale, one might expect a breakdown of the low energy description
at energies $\sim N^{1\over 4}$\cite{Dhar:2005su}. Exact finite $N$ multi-point correlation functions were used in \cite{Dhar:2005su} 
to explore this expectation. Using an operator bosonization of a finite number of nonrelativistic fermions\cite{Dhar:2005fg}, exact 
computations of three-point correlators show that perturbation theory only breaks down at the scale $N^{1\over 2}$\cite{Dhar:2005su}.
There are at least two other arguments for this scale: in \cite{Balasubramanian:2001nh} the groundstate 
wavefunction\cite{Balasubramanian:1998sn} for a scalar field on AdS$_5$ arising from a mode with large angular momentum $l$ was
studied. The size of this wavefunction for an LLM graviton decreases with energy; at an energy scale of $\sim N^{1\over 2}$ it becomes 
of order the Plank scale. An argument for this scale can also be made directly in the field theory: for operators constructed
with a small number of Higgs fields, operators with the same number of fields but with a different number of traces are orthogonal
at large $N$. Identifying the number of traces in an operator with particle number, the supergravity Fock space structure
emerges. However, when the number of fields are of order $N^{1\over 2}$, operators with a different number of traces are no longer
orthogonal\cite{Beisert:2002bb,Constable:2002hw} and ``trace number'' is just not a good quantum number. To go beyond these
energies, one needs to employ a new set of degrees of freedom that are weakly coupled and hence provide a more meaningful
description of the bulk physics than gravitons. In \cite{Dhar:2005su} it was argued that these new degrees of freedom are the 
giant gravitons. The point is that even at high energies, giant gravitons remain weakly interacting. The relevant correlator 
computations are in the half-BPS sector. By using our product rule we can compute giant graviton correlators that go beyond
the half-BPS sector. Our results are consistent with those of \cite{Dhar:2005su} and we find that even in this more general 
setting giant gravitons continue to furnish suitable high energy degrees of freedom. 

Schur polynomials built from a single matrix have, by now, a clear interpretation in the dual gravitational theory. Developing an interpretation
for restricted Schur polynomials in an important problem. Our methods allow a study of the interaction of two restricted Schur polynomials, which 
should shed light on this issue. This is indeed the case: we suggest a physical interpretation for the labels of the restricted
Schur polynomial: the composite operator $\chi_{R,(r_n,r_m)}(Z,X)$ is constructed from the half BPS ``partons'' $\chi_{r_n}(Z)$ and $\chi_{r_m}(X)$. 
Evidence for this proposal comes from considering composite operators in which the constituent partons are weakly interacting: the interaction of 
two such composites is largely determined by the interactions of the partons. 

In the remainder of this introduction we will describe the organization of this paper.
Section 2 describes the product rule satisfied by the restricted Schur polynomials. We focus on polynomials built using
two complex Higgs fields, $X$ and $Z$. The analogous product rule
for the Schur polynomials is determined by the Littlewood-Richardson numbers. For this reason, we call the numbers that
enter our rule {\sl restricted Littlewood-Richardson numbers}. In section 3 we develop techniques that can be used to evaluate
the restricted Littlewood-Richardson numbers. Essentially, we perform a change of basis so that in the new basis we can
compute the restricted characters using the strand diagram techniques developed in \cite{Bekker:2007ea}. These methods are not very efficient
and one is not able, in general, to effectively deal with restricted Schur polynomials which have both a large number of $X$ fields and a
large number of $Z$ fields. There are however special cases for which we can obtain explicit results. These results are used in section 4 
to describe some aspects of the fully quantum mechanical bulk gravity. In particular, we provide evidence that the giant gravitons
(restricted Schur polynomials for the totally antisymmetric representations) furnish suitable high energy degrees of freedom, extending the 
half-BPS studies of \cite{Dhar:2005su}. In section 5 we present the evidence for our proposal that
the composite operator $\chi_{R,(r_n,r_m)}(Z,X)$ is constructed from the half BPS ``partons'' $\chi_{r_n}(Z)$ and $\chi_{r_m}(X)$. 
In section 6 we discuss our results and outline some open problems.

\section{A Product Rule for Restricted Schur Polynomials}

The Schur polynomial $\chi_R(U)$ gives the character of $U\in SU(N)$ in the $SU(N)$ irreducible representation labeled 
by Young diagram $R$. The representation obtained by taking the direct product of two irreducible $SU(N)$ representations
$R_1$ and $R_2$ is in general reducible. The number of times that irreducible representation $T$ appears is given by the
Littlewood-Richardson number $f_{R_1\, R_2;T}$. Using the fact that the Schur polynomials compute characters, it is clear that
$$\chi_{R_1}(Z)\chi_{R_2}(Z)=\sum_T f_{R_1\, R_2 ;T}\chi_T(Z).$$
Using this product rule, it is easy to compute multipoint functions in terms of two point functions. For example, for three point
functions we have
$$\langle \chi_{R_1}(Z)\chi_{R_2}(Z)\chi_S(Z)^\dagger\rangle=\sum_T f_{R_1\, R_2 ;T}\langle\chi_T(Z)\chi_S(Z)^\dagger\rangle\, ,$$
and for four point functions
\begin{eqnarray}
\langle \chi_{R_1}(Z)\chi_{R_2}(Z)\chi_{R_3}(Z)\chi_S(Z)^\dagger\rangle
&=&\sum_T f_{R_1\, R_2 ;T}\langle\chi_T(Z)\chi_{R_3}(Z)\chi_S(Z)^\dagger\rangle
\nonumber\\
&=&\sum_T \sum_U f_{R_1\, R_2 ;T}f_{R_3\, T ;U}\langle\chi_U(Z)\chi_S(Z)^\dagger\rangle\, .
\nonumber
\end{eqnarray}
These results are in perfect agreement with the exact computations of \cite{Corley:2001zk},\cite{Corley:2002mj}. It is also
clear that we can compute $n$-point functions knowing only the product rule and the two point functions.
In this section we will argue that the restricted Schur polynomials themselves satisfy a simple product rule. 

\subsection{Dual Characters}

To begin, we review Appendix I of \cite{de Mello Koch:2007uu}. If two permutations $\sigma,\tau$ satisfy
$$\Tr (\sigma Z^{\otimes n}\otimes X^{\otimes m}) = \Tr (\tau Z^{\otimes n}\otimes X^{\otimes m}),$$
we say they are {\it restricted conjugate}. Restricted conjugate is an equivalence relation. Clearly two elements $\sigma$ and $\tau$
are restricted conjugate if they satisfy
$$ \sigma = \mu^{-1}\tau\mu ,$$
for some $\mu\in S_n\times S_m$. Denote the number of restricted conjugate classes by $N(n,m)$ and let $n^\sigma(n,m)$ denote the number 
of elements in the restricted conjugate class with representative $\sigma$. $N(n,m)$ is also equal to the number of restricted Schur 
polynomials $\chi_{R,(r_n,r_m)}$ with $R$ an irreducible representation of $S_{n+m}$ and $(r_n,r_m)$ an irreducible representation of 
$S_n\times S_m$. The equality between the number of restricted conjugate classes $N(n,m)$ and the total number of labels $R,(r_n,r_m)$ 
generalizes the familiar equality for the symmetric group of the number of conjugacy classes and the number of irreducible representations. 
Introduce the matrix
\begin{equation}
(M^{-1})_{\sigma\tau}=\sum_{(R,(r_n,r_m))}\chi_{R,(r_n,r_m)}(\sigma )\chi_{R,(r_n,r_m)}(\tau),
\label{DefM}
\end{equation}
where the sum on the right hand side runs over all possible labels $(R,(r_n,r_m))$. The restricted character is defined by\cite{de Mello Koch:2007uu}
$$\chi_{R,(r_n,r_m)} (\sigma )=\Tr_{(r_n,r_m)}\left(\Gamma_R(\sigma) \right).$$
Define the {\sl dual restricted character} (which we denote by a superscript) by
$$ \chi^{R,(r_n,r_m)}(\sigma )={n!m!\over n^\sigma(n,m)}\sum_{\big[ \tau\big]_r }M_{\sigma\tau}\chi_{R,(r_n,r_m)}(\tau ).$$
The indices $\sigma$ and $\tau$ that appear in (\ref{DefM}) run over the restricted conjugacy classes. Denote the restricted
conjugacy class with representative $\psi$ by $\big[\psi\big]_r$. It is clear that
\begin{equation}
\sum_{R,(r_n,r_m)} \chi^{R,(r_n,r_m)}(\sigma ) \chi_{R,(r_n,r_m)}(\rho )=n!m!\delta (\big[\sigma\big]_r \big[\rho\big]_r).
\label{delta}
\end{equation}
In appendix A we derive a formula for the dual restricted character in terms of the restricted character. To correctly state this formula, we need to
spell out both the row and the column indices that are summed in the trace. For an ``on the diagonal block'' trace, the column and row indices that are
summed come from the same carrier space and we are summing diagonal elements of the matrix; for an ``off the diagonal block'' trace, the column and row 
indices that are traced come from distinct carrier spaces so that we are summing off diagonal elements of the matrix\footnote{See 
\cite{Balasubramanian:2004nb},\cite{de Mello Koch:2007uu} for more details.}. Indicating both row and column indices of the restricted trace,
the dual character is given by
\begin{equation}
\chi^{R,((r_n,r_m),(s_n,s_m))}(\tau )= {d_R n! m!\over d_{r_{n}}d_{r_{m}}(n+m)!}
\chi_{R,((s_n,s_m),(r_n,r_m))}(\tau ).
\label{dchar}
\end{equation}

\subsection{Product Rule}

We will now define what we call {\it restricted Littlewood-Richardson numbers}. The restricted Littlewood-Richardson numbers determine the product 
rule for restricted Schur polynomials in exactly the same way that the Littlewood-Richardson numbers determine the product rule for Schur polynomials. 
Let $R_1$ be an irreducible representation of $S_{n_1+m_1}$ and let $(r_{n_1},r_{m_1})$ be an irreducible representation of $S_{n_1}\times S_{m_1}$. 
Let $R_2$ be an irreducible representation of $S_{n_2+m_2}$ and let $(r_{n_2},r_{m_2})$ be an irreducible representation of $S_{n_2}\times S_{m_2}$. 
Finally, let $R_{1+2}$ be an irreducible representation of $S_{n_1+n_2+m_1+m_2}$ and let $(r_{n_{1+2}},r_{m_{1+2}})$ be an irreducible representation 
of $S_{n_1+n_2}\times S_{m_1+m_2}$. The restricted Littlewood-Richardson numbers are defined by
\begin{equation} 
f_{R_1,(r_{n_1},r_{m_1})\, R_2,(r_{n_2},r_{m_2})}^{R_{1+2},(r_{n_{1+2}},r_{m_{1+2}})}= {1 \over n_1! n_2! m_1! m_2!}
\label{LRdefn}
\end{equation}
$$\times \sum_{\sigma_1\in S_{n_1+m_1}}\sum_{\sigma_2\in S_{n_2+m_2}} \chi_{R_1,(r_{n_1},r_{m_1})}(\sigma_1)\chi_{R_2,(r_{n_2},r_{m_2})}(\sigma_2)
\chi^{R_{1+2},(r_{n_{1+2}},r_{m_{1+2}})}(\sigma_1\circ\sigma_2).$$
To streamline the notation, from now on we will replace the composite label $R_i,(r_{n_i},r_{m_i})$ simply by $\{ i\}$ and we define
$n_{12}\equiv n_1+n_2$, $m_{12}\equiv m_1+m_2$. With the new streamlined
notation we write
$$ f_{\{ 1\}\, \{ 2\}}^{\{1+2\}}= {1 \over n_1! n_2! m_1! m_2!}
   \sum_{\sigma_1\in S_{n_1+m_1}}\sum_{\sigma_2\in S_{n_2+m_2}} \chi_{\{1\}}(\sigma_1)\chi_{\{2\}}(\sigma_2)
   \chi^{\{1+2\}}(\sigma_1\circ\sigma_2).$$
The restricted Schur product rule says
\begin{equation}
\chi_{\{ 1\}}(Z,X)\chi_{\{ 2\}}(Z,X)=\sum_{\{1+2\}} f_{\{ 1\}\,\{ 2\}}^{\{1+2\}}\chi_{\{1+2\}}(Z,X).
\label{ProdRule}
\end{equation}
A few comments are in order. The restricted Schur polynomial $\chi_{\{1+2\}}(Z,X)$ is given by
$$ \chi_{\{1+2\}}(Z,X)={1\over n_{12}!m_{12}!}
\sum_{\rho\in S_{n_{12}+m_{12}}}
\Tr_{(r_{n_{1+2}},r_{m_{1+2}})}\left(\Gamma_{R_{1+2}}(\rho)\right) \Tr (\rho Z^{\otimes n_{12}}\otimes X^{\otimes m_{12}}).$$
$(r_{n_{1+2}},r_{m_{1+2}})$ is an irreducible representation of the $S_{n_{12}}\times S_{m_{12}}$ subgroup which permutes
indices of the $Z$s amongst each other and the indices of the $X$s amongst each other. The representation $R_1$ is an irreducible
representation of the $S_{n_1+m_1}$ subgroup that acts on the first $n_1$ $Z$s and the first $m_1$ $X$s; the representation $R_2$ 
is an irreducible representation of the $S_{n_2+m_2}$ subgroup that acts on the remaining $n_2$ $Z$s and $m_2$ $X$s. To demonstrate 
the restricted Schur product rule, consider
$$\sum_{\{1+2\}}f_{\{1\}\,\{ 2\}}^{\{1+2\}}\chi_{\{1+2\}}(Z,X)={1\over n_1! n_2! m_1! m_2!}$$
$$\times\sum_{\{1+2\}}\sum_{\sigma_1\in S_{n_1+m_1}}\sum_{\sigma_2\in S_{n_2+m_2}}
\chi_{\{1\}}(\sigma_1)\chi_{\{2\}}(\sigma_2)\chi^{\{1+2\}}(\sigma_1\circ\sigma_2)$$
$$\times{1\over n_{12}!m_{12}!}\sum_{\rho\in S_{n_{12}+m_{12}}}
\Tr_{(r_{n_{1+2}},r_{m_{1+2}})}\left(\Gamma_{R_{1+2}}(\rho)\right) \Tr (\rho Z^{\otimes n_1+n_2}\otimes X^{\otimes m_1+m_2}).$$
After using (\ref{delta}) we obtain
$$\sum_{\{1+2\}}f_{\{1\}\,\{ 2\}}^{\{1+2\}}\chi_{\{1+2\}}(Z,X)={1\over n_1! n_2! m_1! m_2!}
\sum_{\sigma_1\in S_{n_1+m_1}}\sum_{\sigma_2\in S_{n_2+m_2}}\sum_{\rho\in S_{n_{12}+m_{12}}}
\chi_{\{ 1\}}(\sigma_1)\chi_{\{ 2\}}(\sigma_2)\times$$
$$\times \delta (\big[\sigma_1\circ\sigma_2\big]_r\big[\rho\big]_r)
\Tr (\rho Z^{\otimes n_{12}}\otimes X^{\otimes m_{12}})$$
$$={1\over n_1! n_2! m_1! m_2!}
\sum_{\sigma_1\in S_{n_1+m_1}}\sum_{\sigma_2\in S_{n_2+m_2}}
\chi_{\{ 1\}}(\sigma_1)\chi_{\{ 2\}}(\sigma_2)
\Tr (\sigma_1\circ\sigma_2 Z^{\otimes n_{12}}\otimes X^{\otimes m_{12} })$$
$$={1\over n_1! m_1!}
\sum_{\sigma_1\in S_{n_1+m_1}}\chi_{\{1\}}(\sigma_1)
\Tr (\sigma_1 Z^{\otimes n_1}\otimes X^{\otimes m_1})
{1\over n_2! m_2!}\sum_{\sigma_2\in S_{n_2+m_2}}
\chi_{\{2\}}(\sigma_2)\Tr (\sigma_2 Z^{\otimes n_2}\otimes X^{\otimes m_2})$$
$$=\chi_{\{1\}}(Z,X)\chi_{\{2\}}(Z,X),$$
which proves the product rule.

Using the explicit formula for the dual character (\ref{dchar}) leads to the formula
$$ f_{R_1,(r_{n_1},r_{m_1})\, R_2,(r_{n_2},r_{m_2})}^{R_{1+2},(r_{n_{1+2}},r_{m_{1+2}})(s_{n_{1+2}},s_{m_{1+2}})}=
{n_{12}! m_{12}!\over n_1! n_2! m_1! m_2!(n_{12}+m_{12})!}\times$$
$$ {d_{R_{1+2}}\over d_{r_{n_{1+2}}}d_{r_{m_{1+2}}}}
 \sum_{\sigma_1\in S_{n_1+m_1}}\sum_{\sigma_2\in S_{n_2+m_2}} \chi_{R_1,(r_{n_1},r_{m_1})}(\sigma_1)
\chi_{R_2,(r_{n_2},r_{m_2})}(\sigma_2)\times $$
\begin{equation} 
\chi_{R_{1+2},(s_{n_{1+2}},s_{m_{1+2}})(r_{n_{1+2}},r_{m_{1+2}})}(\sigma_1\circ\sigma_2)
\label{LRFormula}
\end{equation}
for the restricted Littlewood-Richardson numbers.

\section{Computation of the Restricted Littlewood-Richardson Numbers}

In this section we will develop rules that will allow us to compute the restricted characters needed to
evaluate the restricted Littlewood Richardson numbers. A general diagrammatic method, {\sl strand diagrams}, to compute
restricted characters in the case that the polynomial is built from a single matrix has been developed in \cite{Bekker:2007ea}.
In this section we would like to develop methods that are powerful enough to allow the computation of restricted characters
for polynomials built out of both $Z$ and $X$. It is enough to compute the characters of two cycles, since 
any element can be decomposed into a product of two cycles. It is precisely this fact that was exploited to build 
the strand diagrams of \cite{Bekker:2007ea}. In the next two sections the character of arbitrary two cycles for an on the diagonal
block restriction and then an off the diagonal block restriction are computed.
Finally, some example computations of restricted Littlewood-Richardson coefficients are discussed.

\subsection{On the diagonal restricted characters of two cycles}

We need two pieces of information: first, we will
introduce three Casimirs that will be particularly useful. Second, we will argue that all characters for two cycles
which do not belong to $S_n\times S_m$ are equal. Taken together, these two facts will allow us to compute the restricted 
character of an arbitrary two cycle. 

Let the first $n$ indices be associated with the $Z$ matrices and the next $m$ indices be associated with the $X$ fields. 
Greek indices run over $\alpha=1,2,...,m+n$. Indices from the start of the alphabet run over $a = 1,2,...,n$. Indices from the 
middle of the alphabet run over $i=n+1,n+2,...,n+m$. The operator
$$ \hat{O}_{n+m}=\sum_{\alpha <\beta =1}^{n+m}\, (\alpha\beta )$$
is a Casimir of $S_{n+m}$. When acting on an irreducible representation described by a Young Diagram $R$ with $r_i$ boxes
in row $i$ and $c_j$ boxes in column $j$ it takes the value
$$ \hat{O}_{n+m}|R\rangle =\big[\sum_{i} r_i(r_i-1)-\sum_{j} c_j(c_j-1)\big]|R\rangle . $$
The operators
$$ \hat{O}_{n}=\sum_{a<b=1}^{n}\, (ab),\qquad \hat{O}_{m}=\sum_{i<j=n+1}^{n+m}\, (ij),$$
are Casimirs of $S_n\times S_m$. Consider the $S_n\times S_m$ irreducible representation $R$, described by Young diagrams
$r_n$ (for the $Z$s) and $r_m$ (for the $X$s). $r_n$ has $r_{1,i}$ boxes in row $i$ and $c_{1,j}$ boxes in column $j$;
$r_m$ has $r_{2,i}$ boxes in row $i$ and $c_{2,j}$ boxes in column $j$. These Casimirs take the values
$$ \hat{O}_{n}|(r_n,r_m)\rangle =\big[\sum_{i} r_{1,i}(r_{1,i}-1)-\sum_{j} c_{1,j}(c_{1,j}-1)\big]|(r_n,r_m)\rangle ,$$
$$ \hat{O}_{m}|(r_n,r_m)\rangle =\big[\sum_{i} r_{2,i}(r_{2,i}-1)-\sum_{j} c_{2,j}(c_{2,j}-1)\big]|(r_n,r_m)\rangle .$$

If a cycle belongs to $S_n \times S_m$ it has the form $(ab)$ or $(ij)$. In this case
$$ \Tr_{(r_n, r_m)}((ab))= \Tr_{r_n}((ab))d_{r_m},\quad {\rm and}\quad \Tr_{(r_n, r_m)} ((ij))=\Tr_{r_m}((ij))d_{r_n}.$$
We can calculate $\Tr_{r_n}((ab))$ and $\Tr_{r_m}((ij))$ using the results of \cite{Bekker:2007ea}.
We will now argue that all characters for two cycles which do not belong to $S_n\times S_m$ are equal. Any two such cycles can be 
related as
\begin{eqnarray}
\Gamma_R\big((aj)\big)&=&\Gamma_R\big((ab)\big)\Gamma_R\big((jl)\big)\Gamma_R\big((bl)\big)\Gamma_R\big((jl)\big)\Gamma_R\big((ab)\big)\nonumber\\
&=&\Gamma_R\big((ab)\big)^{-1}\Gamma_R\big((jl)\big)^{-1}\Gamma_R\big((bl)\big)\Gamma_R\big((jl)\big)\Gamma_R\big((ab)\big).\nonumber
\end{eqnarray}
Thus,
\begin{eqnarray}
\Tr_{(r_n,r_m)}\Big(\Gamma_R\big((aj)\big)\Big)&=&
\Tr_{(r_n,r_m)}\Big(\Gamma_R\big((ab)\big)^{-1}\Gamma_R\big((jl)\big)^{-1}\Gamma_R\big((bl)\big)\Gamma_R\big((jl)\big)\Gamma_R\big((ab)\big)\Big)\nonumber\\
&=& \Tr_{(r_n,r_m)}\Big(\Gamma_{R_\gamma}\big((ab)\big)^{-1}\Gamma_{R_\gamma}\big((jl)\big)^{-1}\Gamma_R\big((bl)\big)
\Gamma_{R_\gamma}\big((jl)\big)\Gamma_{R_\gamma}\big((ab)\big)\Big)\nonumber\\
&=& \Tr_{(r_n,r_m)}\Big(\Gamma_R\big((bl)\big)\Big).\nonumber
\end{eqnarray}
Clearly, all characters for two cycles which do not belong to $S_n\times S_m$ are equal. 

Taken together, these two facts imply that (the cycle $(aj)$ does not belong to $S_n\times S_m$)
\begin{eqnarray}
\Tr_{(r_n,r_m)}\Big(\Gamma_R\big((aj)\big)\Big)&=& {1\over nm}\sum_{i=n+1}^{n+m}\sum_{a=1}^{n}\Tr_{(r_n,r_m)}\Big(\Gamma_R\big((ia)\big)\Big)
\nonumber\\
&=& {1\over nm}\Tr_{(r_n,r_m)}(\hat{O}_{n+m}-\hat{O}_n-\hat{O}_m).
\label{onblock}
\end{eqnarray}
The eigenvalue of the Casimir operator $\hat{O}_{n+m}-\hat{O}_n-\hat{O}_m$ can be written as the sum of the weights of $R$ minus the sum of the weights of
$r_n$ minus the sum of the weights of $r_m$. This observation allows us to derive much simpler formulas for the case that $r_m$ say, contains only a few boxes.
Indeed, we can imagine that $(r_n,r_m)$ was formed by peeling boxes off $R$ to leave $r_n$ and then combining the
peeled boxes to form $r_m$. In this case, the 
eigenvalue of $\hat{O}_{n+m}-\hat{O}_n-\hat{O}_m$ is given by the sum of weights of the boxes peeled off $R$ minus the sum of the weights of the boxes in $r_m$.
We will now give the simplified versions of (\ref{onblock}) for the cases that $m=1,2$ or $3$.

{\vskip 0.5cm}

\noindent
{\sl Restricted character for $m=1$:\,\,\,} In the following formula, $R$ is an irreducible representation of $S_{n+1}$ and $r_n$ is an irreducible 
representation of $S_n$ of dimension $d_{r_n}$. A single box must be removed from $R$ to obtain $r_n$. Denote the weight of the box that must 
be removed by $c_1$. The simplified character formula is
$$ \Tr _{(r_n,{\tiny \yng(1)})}\Big(\Gamma_R \big((aj)\big)\Big) = {c_1-N\over n}d_{r_n}.$$

{\vskip 0.5cm}

\noindent
{\sl Restricted character for $m=2$:\,\,\,} In the following formula, $R$ is an irreducible representation of $S_{n+2}$ and $r_n$ is an irreducible 
representation of $S_n$ of dimension $d_{r_n}$. Two boxes must be removed from $R$ to obtain $r_n$. Denote the weights of the boxes that must 
be removed by $c_1$ and $c_2$. The simplified character formulas are
$$ \Tr _{(r_n,{\tiny \yng(2)})}\Big(\Gamma_R \big((aj)\big)\Big) = {c_1+c_2-2N-1\over 2n}d_{r_n},$$
$$ \Tr _{(r_n,{\tiny \yng(1,1)})}\Big(\Gamma_R \big((aj)\big)\Big) = {c_1+c_2-2N+1\over 2n}d_{r_n}.$$

{\vskip 0.5cm}

\noindent
{\sl Restricted character for $m=3$:\,\,\,} In the following formula, $R$ is an irreducible representation of $S_{n+3}$ and $r_n$ is an irreducible 
representation of $S_n$ of dimension $d_{r_n}$. Three boxes must be removed from $R$ to obtain $r_n$. Denote the weights of the boxes that must 
be removed by $c_1$, $c_2$ and $c_3$. The simplified character formulas are
$$ \Tr _{(r_n,{\tiny \yng(3)})}\Big(\Gamma_R \big((aj)\big)\Big) = {c_1+c_2+c_3-3N-3\over 3n}d_{r_n},$$
$$ \Tr _{(r_n,{\tiny \yng(2,1)})}\Big(\Gamma_R \big((aj)\big)\Big) = 2{c_1+c_2+c_3-3N\over 3n}d_{r_n},$$
$$ \Tr _{(r_n,{\tiny \yng(1,1,1)})}\Big(\Gamma_R \big((aj)\big)\Big) = {c_1+c_2+c_3-3N+3\over 3n}d_{r_n}.$$

\subsection{Off the diagonal restricted characters of two cycles}

Before we can compute generic restricted characters, we need to compute traces over off the diagonal blocks:
$\Tr_{(r_n,r_m),(s_n,s_m)}\Big(\Gamma_R\big( (\alpha\beta)\big)\Big)$, where $(r_n,r_m)$ and $(s_n,s_m)$ are
distinct $S_{n}\times S_{m}$ representations. This character clearly vanishes if $(\alpha\beta)$ belongs 
to the $S_n\times S_m$ subgroup. What about the $mn$ two cycles that do not belong to $S_n\times S_m$?
For concreteness, consider the computation of $\Tr_{(r_n,r_m)(s_n,s_m)}\Big(\Gamma_R\big( (ai)\big)\Big)$.
Let $(r_n,r_m)_I'$ denote the complete set of irreducible representations of the $S_{n-1}\times S_{m-1}$ subgroup subduced by
$(r_n,r_m)$, that is $\oplus_I\, (r_n,r_m)_I' =(r_n,r_m)$. The $S_{n-1}\times S_{m-1}$ subgroup 
of interest is obtained by keeping all elements of $S_n\times S_m$ that hold indices $a$ and $i$ fixed. 
The representations $(r_n,r_m)$ and $(s_n,s_m)$ have the same shape\footnote{It is only when $(r_n,r_m)$ 
and $(s_n,s_m)$ have the same shape that the trace $\Tr_{(r_n,r_m)(s_n,s_m)}$ has any meaning.} 
so that we can establish a bijective map
between their bases. We will assume that this bijective map is the identity, which we can always arrange by a 
suitable choice of basis. This choice of basis ensures that when we subduce to the $S_{n-1}\times S_{m-1}$ 
subgroup we have
$$ \Tr_{(r_n,r_m)(s_n,s_m)}\Big(\Gamma_R\big( (ai)\big)\Big)=\sum_I
\Tr_{(r_n,r_m)_I' (s_n,s_m)_I'}\Big(\Gamma_R\big( (ai)\big)\Big).$$
We will now provide further insight into this formula.
It is straight forward to prove that
$$ \langle (r_n,r_m)_I',i |\Gamma_R\big( (ai)\big)\Big) |(s_n,s_m)_I'\rangle $$
vanishes unless $(r_n,r_m)_I'$ and $(s_n,s_m)_I'$ have the same shape. 
Introduce the Casimirs
$$ \hat{O}_{n-1}=\sum_{c<b=1\ne a}^{n}\, (cb),\qquad \hat{O}_{m-1}=\sum_{k<j=n+1\ne i}^{n+m}\, (kj).$$
Denote the eigenvalue of Casimir $\hat{O}_{n-1}$ for the $(r_n,r_m)_I'$ representation by $\lambda^{(n-1)}_{r}$
and the eigenvalue of Casimir $\hat{O}_{m-1}$ for the $(r_n,r_m)_I'$ representation by $\lambda^{(m-1)}_{r}$.
Clearly, we have
\begin{eqnarray}
\nonumber
\lambda^{(n-1)}_{r}\langle (r_n,r_m)_I',i |\Gamma_R\big( (ai)\big)|(s_n,s_m)_I'\rangle
&=&\langle (r_n,r_m)_I',i |\hat{O}_{n-1}\Gamma_R\big( (ai)\big)|(s_n,s_m)_I'\rangle\\
&=&\langle (r_n,r_m)_I',i |\Gamma_R\big( (ai)\big)\hat{O}_{n-1}|(s_n,s_m)_I'\rangle\nonumber\\
&=&\lambda^{(n-1)}_s\langle (r_n,r_m)_I',i |\Gamma_R\big( (ai)\big)|(s_n,s_m)_I'\rangle,\nonumber
\end{eqnarray}
\begin{eqnarray}
\nonumber
\lambda^{(m-1)}_{r}\langle (r_n,r_m)_I',i |\Gamma_R\big( (ai)\big)|(s_n,s_m)_I'\rangle
&=&\langle (r_n,r_m)_I',i |\hat{O}_{m-1}\Gamma_R\big( (ai)\big)|(s_n,s_m)_I'\rangle\\
&=&\langle (r_n,r_m)_I',i |\Gamma_R\big( (ai)\big)\hat{O}_{m-1}|(s_n,s_m)_I'\rangle\nonumber\\
&=&\lambda^{(m-1)}_{s}\langle (r_n,r_m)_I',i |\Gamma_R\big( (ai)\big)|(s_n,s_m)_I'\rangle,\nonumber
\end{eqnarray}
which implies that
$$\langle (r_n,r_m)_I',i |\Gamma_R\big( (ai)\big)|(s_n,s_m)_I'\rangle\propto
\delta_{\lambda^{(n-1)}_{r}\, \lambda^{(n-1)}_{s}}
\delta_{\lambda^{(m-1)}_{r}\, \lambda^{(m-1)}_{s}}.$$
We can repeat this with the complete set of Casimirs of $S_{n-1}\times S_{m-1}$, allowing us to conclude that
$$\langle (r_n,r_m)_I',i |\Gamma_R\big( (ai)\big)|(s_n,s_m)_I'\rangle\propto
\delta_{(r_n,r_m)_I' \, (s_n,s_m)_I'} .$$
$\delta_{(r_n,r_m)_I' \, (s_n,s_m)_I'}$ is equal to 1 if $(r_n,r_m)_I'$ and $(s_n,s_m)_I'$ have the same shape. Of course,
this is just a consequence of Schur's Lemma: since $\Gamma_R\big( (ai)\big)$ commutes
with all of the elements of the $S_{n-1}\times S_{m-1}$ subgroup, it is proportional to the identity when acting on any irreducible
representation of $S_{n-1}\times S_{m-1}$. Thus, (the label $i$ inside the ket labels the carrier space state)
$$\langle (r_n,r_m)_I',j| \Gamma_R\big( (ai)\big)|(r_n,r_m)_I',i\rangle = \eta_{(r_n,r_m)_I'}\delta_{ij}\, .$$
If we allow $\Gamma_R\big( (ai)\big)$ to act in the full carrier space of $R$, this equation is modified to
(the sum is over all irreducible representations $(t_n,t_m)$ that can subduce a $(t_n,t_m)'_K$ of the same shape
as $(r_n,r_m)_I'$; the $\delta_{(r_n,r_m)_I' \, (s_n,s_m)_J'}$ on the last line is there to remind the reader
that a non-zero result is obtained only if $(r_n,r_m)_I'$ and  $(s_n,s_m)_J'$ have the same shape)
\begin{eqnarray}
\langle (s_n,s_m)_J',j |\Gamma_R\big( (ai)\big)|(r_n,r_m)_I',i\rangle &=& \sum_{(t_n,t_m)'_K}
\eta_{(r_n,r_m)_I'}^{(t_n,t_m)_K'}\langle (s_n,s_m)_J',j|(t_n,t_m)_K',i\rangle
\nonumber\\
&=&  \eta_{(r_n,r_m)_I'}^{(s_n,s_m)_J'} \delta_{(r_n,r_m)_I' \, (s_n,s_m)_J'}\delta_{ij}\nonumber
\end{eqnarray}
Although $(s_n,s_m)_J'$ and $(r_n,r_m)_I'$ have the same shape, they may be distinct representations in which case they were 
subduced by {\sl different} $S_n\times S_m$ representations. These matrix elements are needed to provide a complete generalization
of strand diagrams. This has an important implication: in the present case, the most general non-vanishing strand diagrams will not 
only record the reordering of boxes in the row and column states, it will also allow the shape of the row state $S_n\times S_m$
representation to reorder itself (by the movement of a single box in $r_n$ and/or a single box in $r_m$) into the column state $S_n\times S_m$
representation. 

The trace we are interested in can also be expressed in terms of $\eta_{(r_n,r_m)_J'}^{(r_n,r_m)_I'}$. Indeed, it is clear that
\begin{eqnarray} 
\Tr_{(r_n,r_m)(s_n,s_m)}\Big(\Gamma_R\big( (ai)\big)\Big)&=&\sum_I
\Tr_{(r_n,r_m)_I' (s_n,s_m)_I'}\Big(\Gamma_R\big( (ai)\big)\Big)\nonumber\\
&=&\sum_{I} \eta_{(r_n,r_m)_I'}^{(r_n,r_m)_I'}d_{(r_n,r_m)_I'}.\nonumber
\end{eqnarray}
$d_{(r_n,r_m)_I'}$ is the dimension of the irreducible representation $(r_n,r_m)_I'$. Since $(r_n,r_m)$ and $(s_n,s_m)$ have the same shape,
we will not need to worry about the extra complication of changing the $S_n\times S_m$ representation between the row and column states.

To determine the restricted character we need, we now only need to fix the $\eta_{(r_n,r_m)_I'}^{(r_n,r_m)_I'}$. The most direct way we 
have found to do this proceeds by determining the explicit change of basis from a natural basis of
$S_{n+m}$ to a natural basis for the $S_n\times S_m$ subgroup. We will illustrate the method with an example: 
using the methods of the last subsection, we easily find
$$\Tr_{\tiny \yng(2,1),\yng(2)}\left(\Gamma_{\tiny \yng(3,1,1)}\Big((3,4)\Big)\right)=-{1\over 3}.$$
To extract ${\tiny \yng(2,1)}$ from ${\tiny \yng(3,1,1)}$ we need to pull off the last box in the first row and the last box in
the first column. They can be pulled off in any order, so that we can write ($i$ labels the states in the ${\tiny \yng(2,1)}$ carrier
space)
\begin{equation} 
|{\tiny \yng(3,1,1)};{\tiny \yng(2,1)}\, i, {\tiny\yng(2)}\rangle =\alpha |{\tiny\young({\,}{\,}{1},{\,},{2})}\, i\rangle
+ \beta |{\tiny\young({\,}{\,}{2},{\,},{1})}\, i\rangle .
\label{changebasis}
\end{equation}
Because the above state is normalized, we need $\alpha^2 + \beta^2 = 1$. Next, using appendix D.2 of  \cite{de Mello Koch:2007uu} we can write
$$ \Gamma_{\tiny \yng(3,1,1)}\left( (45)\right)|{\tiny \yng(3,1,1)};{\tiny \yng(2,1)}\, i, {\tiny\yng(2)}\rangle =
|{\tiny \yng(3,1,1)};{\tiny \yng(2,1)}\, i, {\tiny\yng(2)}\rangle,$$
$$ \Gamma_{\tiny \yng(3,1,1)}\left( (45)\right)\left[
\alpha |{\tiny\young({\,}{\,}{1},{\,},{2})}\, i\rangle
+ \beta |{\tiny\young({\,}{\,}{2},{\,},{1})}\, i\rangle\right]=
\left({\alpha\over 4}+\beta{\sqrt{15}\over 4}\right) |{\tiny\young({\,}{\,}{1},{\,},{2})}\, i\rangle
+ \left(-{\beta\over 4}+\alpha{\sqrt{15}\over 4}\right) |{\tiny\young({\,}{\,}{2},{\,},{1})}\, i\rangle .$$
This then implies two equations
$${\alpha\over 4}+\beta{\sqrt{15}\over 4}=\alpha,\qquad -{\beta\over 4}+\alpha{\sqrt{15}\over 4}=\beta .$$
They are not independent (as expected) and imply $\alpha=\sqrt{5\over 3}\beta$ so that
$$\alpha =\sqrt{5\over 8},\qquad \beta=\sqrt{3\over 8}.$$
It is now straight forward to use the standard strand diagram techniques of \cite{Bekker:2007ea} to verify that
$$ \sum_i \left( \sqrt{5\over 8} \langle {\tiny\young({\,}{\,}{1},{\,},{2})}\, i |
+ \sqrt{3\over 8} \langle {\tiny\young({\,}{\,}{2},{\,},{1})}\, i |\right)
\Gamma_{\tiny \yng(3,1,1)}\Big((3,4)\Big)
\left( \sqrt{5\over 8} |{\tiny\young({\,}{\,}{1},{\,},{2})}\, i\rangle
+ \sqrt{3\over 8} |{\tiny\young({\,}{\,}{2},{\,},{1})}\, i\rangle\right)=-{1\over 3}.$$
Although we have computed an on the diagonal character, it is clear that once the relationship (\ref{changebasis}) 
is established, it can be used to compute off the diagonal block characters by employing standard strand diagrams. 

{\vskip 0.25cm}

\noindent
{\bf Summary of the logic:} Using the action of 2-cycles from the $S_m$ subgroup - something we already know in both bases - we have been able to determine
the explicit change of basis from a natural basis of $S_{n+m}$ to a natural basis for the $S_n\times S_m$ subgroup.
This has enabled us to compute the restricted character of the two cycle $(n,n+1)$ which ``straddles'' $S_n$ and
$S_m$. Thus, we can now compute the restricted characters of all two cycles of the form $(i,i+1)$, which is all that
is needed to compute the restricted character of a general group element.

{\vskip 0.5cm}

We will now argue that, as long as the number of boxes in $r_m$ is small, we can find formulas for the above change of basis
for any representation $r_n$. Removing the $m$ boxes (used to build $r_m$) from the $S_{n+m}$ representation $R$ gives a set
of states that carry an index for the carrier space of $r_n$. The coefficient describing the change of basis is clearly 
independent of this index. Thus, in what follows, the carrier space index of representation $r_n$ plays no role and is hence 
suppressed. 

{\vskip 0.5cm}

\noindent
{\sl Restricted character of $(n,n+1)$ for $m=1$:\,\,\,} Denote the two labels of the restricted trace by $R,(r_n,{\tiny\yng(1)})$
and $R,(s_n,{\tiny\yng(1)})$. $R$ is a Young diagram with $n+1$ boxes; $r_n$ and $s_n$ are both Young diagrams with $n$ boxes;
they are both obtained by removing a single box from $R$. Denote the weight of the box that must be removed from $R$ to obtain
$r_n$ by $c_r$; denote the weight of the box that must be removed from $R$ to obtain $s_n$ by $c_s$. For an off the diagonal block, 
$c_r\ne c_s$. Let $R''$ denote the Young diagram obtained when both boxes are removed. It is now straight forward to show that
$$ \Tr_{(r_n,{\tiny \yng(1)})(s_n,{\tiny \yng(1)})}\left(\Gamma_{R}\Big( (n,n+1)\Big)\right)=\sqrt{1-{1\over (c_r-c_s)^2}}d_{R''},$$
where $d_{R''}$ is the dimension of $R''$.

{\vskip 0.5cm}

\noindent
{\sl Restricted character of $(n,n+1)$ for $m=2$:\,\,\,} The two possible types of restricted Schur labels are $R,(r_n,{\tiny\yng(2)})$
and $R,(r_n,{\tiny\yng(1,1)})$. $R$ is a Young diagram with $n+2$ boxes; $r_n$ is a Young diagram with $n$ boxes. $r_n$ is obtained from
$R$ by removing two boxes from $R$. We say that box $a$ is above box $b$ if the weight of box $a$ is greater than the weight of box $b$. Denote the weights
of the two boxes as $c_1$ and $c_2$, such that the box with weight $c_1$ is above the box with weight $c_2$. The relations between the
natural $S_{n+2}$ basis and the natural $S_n\times S_2$ bases are (although we have written these relations using a particular Young diagram,
they are true in general; on the right hand side, the box with label $a$ is to be removed first)
$$ |{\tiny \yng(3,1,1)};{\tiny \yng(2,1)}\, i, {\tiny\yng(2)}\rangle =
 \sqrt{c_1-c_2+1\over 2(c_1-c_2)}
|{\tiny\young({\,}{\,}{a},{\,},{b})}\, i\rangle
+ \sqrt{c_1-c_2-1 \over 2(c_1-c_2)} |{\tiny\young({\,}{\,}{b},{\,},{a})}\, i\rangle ,$$
$$ |{\tiny \yng(3,1,1)};{\tiny \yng(2,1)}\, i, {\tiny\yng(1,1)}\rangle =
\sqrt{c_1-c_2-1 \over 2(c_1-c_2)}|{\tiny\young({\,}{\,}{a},{\,},{b})}\, i\rangle
-\sqrt{c_1-c_2+1\over 2(c_1-c_2)}|{\tiny\young({\,}{\,}{b},{\,},{a})}\, i\rangle .$$
Notice that these states are orthogonal as they must be. It is now straight forward to obtain any particular character
we want by employing standard strand diagram methods. 

We will now consider a particular example. The labels for the restricted trace are $R,(r_n,{\tiny\yng(2)})$ and $R,(s_n,{\tiny\yng(2)})$.
To obtain a non-zero off the diagonal restricted character, one of the boxes removed from $R$ to obtain $r_n$ must be in the same position
as one of the boxes removed from $R$ to obtain $s_n$. Assume that the common box has weight $c_1$. Denote the weight of the second box that
must be removed to obtain $r_n$ by $c_2$ and denote the weight of the second box that must be removed to obtain $s_n$ by $c_2^*$. Denote the
Young diagram obtained by removing all three boxes from $R$ by $R'''$. It is straight forward to show that
$$ \Tr_{(r_n,{\tiny \yng(2)})(s_n,{\tiny \yng(2)})}\left(\Gamma_{R}\Big( (n,n+1)\Big)\right)=
\sqrt{c_1-c_2+1\over 2(c_1-c_2)}\sqrt{c_1-c_2^*+1\over 2(c_1-c_2^*)}\sqrt{1-{1\over (c_2-c_2^*)^2}}d_{R'''}.$$

{\vskip 0.5cm}

\noindent
{\sl Restricted character of $(n,n+1)$ for $m=3$:\,\,\,} We will discuss this example in some detail because it will provide
the key to obtaining general results. We will consider the case in which the three boxes to be removed have no sides in common.
In this case, the three possible types of restricted Schur labels are $R,(r_n,{\tiny\yng(3)})$, $R,(r_n,{\tiny\yng(2,1)})$
and $R,(r_n,{\tiny\yng(1,1,1)})$. $R$ is a Young diagram with $n+3$ boxes; $r_n$ is a Young diagram with $n$ boxes. $r_n$ is obtained from
$R$ by removing three boxes from $R$. Denote the weights of the boxes to be removed by $c_1$, $c_2$ and $c_3$. The box with weight $c_1$ lies
above the boxes with weights $c_2$ and $c_3$; the box with weight $c_2$ lies above the box with weight $c_3$. Consider the expansion
(on the right hand side, the box with label $a$ is to be removed first the box with label $b$ second and the box with label $c$ third)
\begin{eqnarray} 
|{\tiny \yng(3,2,1) };{\tiny \yng(2,1)\,\yng(3)}\rangle
=   \alpha_{123}|{\tiny \young({\,}{\,}{a},{\,}{b},{c})}\rangle
&+& \alpha_{132}|{\tiny \young({\,}{\,}{a},{\,}{c},{b})}\rangle
+   \alpha_{213}|{\tiny \young({\,}{\,}{b},{\,}{a},{c})}\rangle\nonumber\\
+   \alpha_{231}|{\tiny \young({\,}{\,}{b},{\,}{c},{a})}\rangle
&+& \alpha_{312}|{\tiny \young({\,}{\,}{c},{\,}{a},{b})}\rangle
+   \alpha_{321}|{\tiny \young({\,}{\,}{c},{\,}{b},{a})}\rangle\, .
\nonumber
\end{eqnarray}
Notice that the subscripts encode the order in which boxes are to be dropped. By considering the action of the cycle of $(n+2,n+3)$ on the
above expression, we obtain 6 equations. Only three of these are independent; they read
$$ \alpha_{123}={1\over c_1-c_2}\alpha_{123}+\sqrt{1-{1\over (c_1-c_2)^2}}\alpha_{213}, $$
$$ \alpha_{132}={1\over c_1-c_3}\alpha_{132}+\sqrt{1-{1\over (c_1-c_3)^2}}\alpha_{231}, $$
$$ \alpha_{321}={1\over c_3-c_2}\alpha_{321}+\sqrt{1-{1\over (c_3-c_2)^2}}\alpha_{312}. $$
Similarly, by considering the action of the cycle $(n+1,n+2)$ we obtain the following three independent equations
$$ \alpha_{123}={1\over c_2-c_3}\alpha_{123}+\sqrt{1-{1\over (c_2-c_3)^2}}\alpha_{132}, $$
$$ \alpha_{213}={1\over c_1-c_3}\alpha_{213}+\sqrt{1-{1\over (c_1-c_3)^2}}\alpha_{312}, $$
$$ \alpha_{321}={1\over c_2-c_1}\alpha_{321}+\sqrt{1-{1\over (c_2-c_1)^2}}\alpha_{231}. $$
These six equations can be written in a very compact form: let $p$ denote the subscript of a particular coefficient, i.e. a particular ordering
of the three numbers $1$, $2$ and $3$. Define the action of the cycle $(i,i+1)$ on $p$ as follows: all numbers not equal to $i$ or $i+1$  
stay where they are; the numbers $i$ and $i+1$ swap positions. Thus, $(1,2)\,\cdot \,123 = 213$. We will also index the entries of $p$ 
by $i=1,2,3$. Thus, for $p=231$, we have $p(1)=2,$ $p(2)=3$ and $p(3)=1$. The six equations above can now be written as
\begin{equation}
\alpha_p={1\over c_{p^{-1}(i)}-c_{p^{-1}(i+1)}}\alpha_p+\sqrt{1-{1\over (c_{p^{-1}(i)}-c_{p^{-1}(i+1)})^2}}\alpha_{(i,i+1)\,\cdot\, p},\qquad i=1,2,...,m-1.
\label{general}
\end{equation}
Simple algebra gives
$$
\alpha_p=\sqrt{c_{p^{-1}(i)}-c_{p^{-1}(i+1)}+1\over c_{p^{-1}(i)}-c_{p^{-1}(i+1)}-1}\alpha_{(i,i+1)\,\cdot\, p}\, .
$$
These equations completely determine the unknown coefficients. Indeed, each subscript is a particular ordering of the three numbers $1$, $2$ and $3$.
By using the first set of equations we can swap the positions of $1$ and $2$; by using the second set we can swap the positions of $2$ and $3$. Using
these two operations we can relate any coefficient to any other. Thus, the normalization condition
$$ \alpha_{123}^2+\alpha_{132}^2+\alpha_{213}^2+\alpha_{231}^2+\alpha_{312}^2+\alpha_{321}^2 = 1$$
can be written completely in terms of $\alpha_{123}$ (say). This determines $\alpha_{123}$ and hence all of the coefficients. Solving for $\alpha_{123}$
we obtain
$$\alpha_{123}=\sqrt{(c_1-c_2+1)(c_1-c_3+1)(c_2-c_3+1)\over 6(c_1-c_2)(c_1-c_3)(c_2-c_3)}.$$

These conclusions are rather general: the equations following from applications of two cycles of the form $(n+i,n+i+1)$, $i=1,2,...,m-1$ are given
by (\ref{general}) for any state on the right hand side with label $R,(r_n,(m)\, )$ where $(m)$ is the completely symmetric representation.
For any state on the right hand side, the equations following from application of the two cycles $(i,i+1)$, $i=1,2,...,m-1$ together with
the normalization condition determines the expansion coefficients uniquely. The resulting solution is
$$\alpha_{123\cdots m}=\sqrt{{1\over m!}\prod_{i=1}^{m-1}\left[\prod_{j=i+1}^{m} {c_i-c_j+1 \over c_i-c_j}\right]}.$$
It is now straight forward to compute restricted characters for any restricted traces with labels $R,(r_n,{\tiny\yng(3)})$.

Next, consider restricted traces with label $R,(r_n,{\tiny\yng(1,1,1)})$. Things work exactly as for the case just considered. The relevant 
expansion is
\begin{eqnarray} 
|{\tiny \yng(3,2,1) };{\tiny \yng(2,1)\,\yng(1,1,1)}\rangle
=   \beta_{123}|{\tiny \young({\,}{\,}{a},{\,}{b},{c})}\rangle
&+& \beta_{132}|{\tiny \young({\,}{\,}{a},{\,}{c},{b})}\rangle
+   \beta_{213}|{\tiny \young({\,}{\,}{b},{\,}{a},{c})}\rangle\nonumber\\
+   \beta_{231}|{\tiny \young({\,}{\,}{b},{\,}{c},{a})}\rangle
&+& \beta_{312}|{\tiny \young({\,}{\,}{c},{\,}{a},{b})}\rangle
+   \beta_{321}|{\tiny \young({\,}{\,}{c},{\,}{b},{a})}\rangle\, .
\nonumber
\end{eqnarray}
We will write the results of our analysis for general $m$.
It is straight forward to obtain
\begin{equation}
-\beta_p={1\over c_{p^{-1}(i)}-c_{p^{-1}(i+1)}}\beta_p+\sqrt{1-{1\over (c_{p^{-1}(i)}-c_{p^{-1}(i+1)})^2}}\beta_{(i,i+1)\,\cdot\, p},\qquad i=1,2,...,m-1,
\label{generall}
\end{equation}
and hence
$$
\beta_p=-\sqrt{c_{p^{-1}(i)}-c_{p^{-1}(i+1)}-1\over c_{p^{-1}(i)}-c_{p^{-1}(i+1)}+1}\beta_{(i,i+1)\,\cdot\, p}.
$$
Solving these equations together with the normalization condition gives
$$\beta_{123\cdots m}\sqrt{{1\over m!}\prod_{i=1}^{m-1}\left[\prod_{j=i+1}^{m} {c_i-c_j-1 \over c_i-c_j}\right]}.$$
As a check of our phase conventions, we have checked that $\alpha_p$ and $\beta_p$ define orthogonal vectors.

We have not been able to treat the last case, restricted traces with labels $R,(r_n,{\tiny\yng(2,1)})$, for general $r_m$. 
Further, the solution we have obtained
is not unique, because the representation ${\tiny \yng(2,1)}$ appears twice in the outer product ${\tiny \yng(1)\times\yng(1)\times\yng(1)}$. We
have tried to use the freedom we have to choose the simplest possible solution. Consider the expansion
\begin{eqnarray} 
|{\tiny \yng(3,2,1) };{\tiny \yng(2,1)\,\yng(2,1)\, i}\rangle^a
=   \gamma^{a}_{123\, i}|{\tiny \young({\,}{\,}{a},{\,}{b},{c})}\rangle
&+& \gamma^{a}_{132\, i}|{\tiny \young({\,}{\,}{a},{\,}{c},{b})}\rangle
+   \gamma^{a}_{213\, i}|{\tiny \young({\,}{\,}{b},{\,}{a},{c})}\rangle\nonumber\\
+   \gamma^{a}_{231\, i}|{\tiny \young({\,}{\,}{b},{\,}{c},{a})}\rangle
&+& \gamma^{a}_{312\, i}|{\tiny \young({\,}{\,}{c},{\,}{a},{b})}\rangle
+   \gamma^{a}_{321\, i}|{\tiny \young({\,}{\,}{c},{\,}{b},{a})}\rangle\, .
\nonumber
\end{eqnarray}
$a$ is a multiplicity label - it takes the values $1,2$. $i$ is a label for states in the carrier space of $r_m$; it takes one of the two values
$$ i=1\leftrightarrow \young(32,1),\qquad i=2\leftrightarrow \young(31,2). $$
A straight forward (but tedious) computation now determines
%
%
$$\gamma^{1}_{123\, 1}=-{(c_{12}-2)\sqrt{(c_{23}+1)(c_{23}-1)(c_{13}+1)}\over d},\qquad 
\gamma^{1}_{132\, 1}=\sqrt{c_{23}-1\over c_{23}+1}\gamma^{1}_{123\, 1},$$
$$\gamma^{1}_{213\, 1}=2{\sqrt{(c_{12}-1)(c_{12}+1)(c_{23}+1)(c_{23}-1)(c_{13}+1)}\over d},\qquad 
\gamma^{1}_{312\, 1}=\sqrt{c_{13}-1\over c_{13}+1}\gamma^{1}_{213\, 1},$$
$$\gamma^{1}_{231\, 1}=-{(c_{12}+1)(c_{23}+2)\sqrt{(c_{13}-1)}\over d},\qquad 
\gamma^{1}_{321\, 1}=\sqrt{c_{12}-1\over c_{12}+1}\gamma^{1}_{231\, 1},$$
%
%
$$\gamma^{1}_{123\, 2}={c_{12}\sqrt{3(c_{23}+1)(c_{23}-1)(c_{13}+1)}\over d},\qquad 
\gamma^{1}_{132\, 2}=-\sqrt{c_{23}+1\over c_{23}-1}\gamma^{1}_{123\, 2},$$
$$\gamma^{1}_{213\, 2}=0,\qquad 
\gamma^{1}_{312\, 2}=0,$$
$$\gamma^{1}_{231\, 2}=-{(c_{12}-1)c_{23}\sqrt{3(c_{13}-1)}\over d},\qquad 
\gamma^{1}_{321\, 2}=-\sqrt{c_{12}+1\over c_{12}-1}\gamma^{1}_{231\, 2},$$
%
%
$$\gamma^{2}_{123\, 1}={(c_{23}+1)\sqrt{3(c_{12}+1)(c_{12}-1)(c_{13}-1)}\over d},\qquad 
\gamma^{2}_{132\, 1}=\sqrt{c_{23}-1\over c_{23}+1}\gamma^{2}_{123\, 1},$$
$$\gamma^{2}_{213\, 1}=-{(c_{13}+1)\sqrt{3(c_{13}-1)}\over d},\qquad 
\gamma^{2}_{312\, 1}=\sqrt{c_{13}-1\over c_{13}+1}\gamma^{2}_{213\, 1},$$
$$\gamma^{2}_{231\, 1}=-{\sqrt{3(c_{12}+1)(c_{12}-1)(c_{23}+1)(c_{23}-1)(c_{13}+1)}\over d},\qquad 
\gamma^{2}_{321\, 1}=\sqrt{c_{12}-1\over c_{12}+1}\gamma^{2}_{231\, 1},$$
%
%
$$\gamma^{2}_{123\, 2}={(c_{23}-1)\sqrt{(c_{12}+1)(c_{12}-1)(c_{13}-1)}\over d},\qquad 
\gamma^{2}_{132\, 2}=-\sqrt{c_{23}+1\over c_{23}-1}\gamma^{2}_{123\, 2},$$
$$\gamma^{2}_{213\, 2}={(2c_{12}c_{23}+c_{12}-c_{23}+1)\sqrt{c_{13}-1}\over d},\qquad 
\gamma^{2}_{312\, 2}=-\sqrt{c_{13}+1\over c_{13}-1}\gamma^{2}_{213\, 2},$$
$$\gamma^{2}_{231\, 2}={\sqrt{(c_{12}+1)(c_{12}-1)(c_{23}+1)(c_{23}-1)(c_{13}+1)}\over d},\qquad 
\gamma^{2}_{321\, 2}=-\sqrt{c_{12}+1\over c_{12}-1}\gamma^{2}_{231\, 2},$$
where $c_{ij}\equiv c_i-c_j$ and
$$ d=\sqrt{6(c_1-c_2)(c_2-c_3)(c_1-c_3)(2(c_1-c_2)(c_2-c_3)-2c_2+c_3+c_1+1)}.$$
As a partial check of our phase conventions we have verified that
$$ \{ |{\tiny \yng(3,2,1) };{\tiny \yng(2,1)\,\yng(3)}\rangle \, ,\,
|{\tiny \yng(3,2,1) };{\tiny \yng(2,1)\,\yng(1,1,1)}\rangle \, , \,
|{\tiny \yng(3,2,1) };{\tiny \yng(2,1)\,\yng(2,1)\, i}\rangle^a \}$$
provides an orthonormal basis. This does not yet provide a complete treatment of the $m=3$ case, because we have not yet
considered the case that the three boxes removed share common sides. The generalization to this case is straight forward,
using the methods we developed in this section.

\subsection{Some example computations of restricted Littlewood-Richardson Coefficients}

In this subsection we will discuss the computation of the restricted Littlewood-Richardson coefficients that will be needed to study
graviton interactions in the next section. The first case we wish to consider are representations where $R$, $r_n$ and $r_m$ are totally
antisymmetric. In the case of a single matrix, these correlators are dual to branes that have expanded in the S$^5$ of the AdS$_5\times$S$^5$
geometry. Since $n+m$ is cut off at $N$ it is natural to conjecture that the same interpretation holds even when the operator is built using
two complex Higgs fields. As an example, for $n=2$ and $m=3$ we are discussing the restricted representation with labels
$$ R={\tiny \yng(1,1,1,1,1)}\equiv (1^{5}),\qquad (r_n,r_m)=({\tiny \yng(1,1,1)},{\tiny \yng(1,1)})\equiv (1^3,1^2).$$
This particular class of restricted characters are particularly easy to compute: because the dimension of the representation
and the dimensions of the restriction are equal, the restriction is trivial and we have ($\epsilon(\sigma)$ is 1 if $\sigma$ is 
an odd permutation and 0 if $\sigma$ is an even permutation)
$$ \chi_{(1^{n+m}),(1^n,1^m)}(\sigma)=  \chi_{(1^{n+m})}(\sigma)=(-1)^{\epsilon(\sigma)},$$
and
$$ \chi^{(1^{n+m}),(1^n,1^m)}(\sigma )={n!m!\over (n+m)!}\chi_{1^{n+m}}(\sigma ). $$
Consequently, for the case under consideration, we can express the restricted Littlewood-Richardson numbers in terms of the Littlewood-Richardson
numbers as
$$ f^{1^{n_1+n_2+m_1+m_2},(1^{n_1+n_2},1^{m_1+m_2})}_{1^{n_1+m_1},(1^{n_1},1^{m_1})\, 1^{n_2+m_2},(1^{n_2},1^{m_2})}
={n_{12}!m_{12}!(n_1+m_1)!(n_2+m_2)!\over n_1!n_2!m_1!m_2!(n_{12}+m_{12})!}f_{1^{n_1+m_1}\, 1^{n_2+m_2}\, 1^{n_1+n_2+m_1+m_2}},$$
where we have used the formula
$$ f_{R\, S\, T}={1\over n_R! n_S!}\sum_{\alpha_1\in S_R}\sum_{\alpha_2\in S_S}\chi_R(\sigma_1)\chi_S(\sigma_2)\chi_T(\sigma_1\circ\sigma_2).$$
It is satisfying that upon setting $m_1=m_2=0$ our restricted Littlewood-Richardson numbers reproduce the standard Littlewood-Richardson numbers.
Rules for the computation of the Littlewood-Richardson coefficients can be found, for example, in \cite{FH}. For the case under consideration
here, the Littlewood-Richardson number is 1, so that
$$ f^{1^{n_1+n_2+m_1+m_2},(1^{n_1+n_2},1^{m_1+m_2})}_{1^{n_1+m_1},(1^{n_1},1^{m_1})\, 1^{n_2+m_2},(1^{n_2},1^{m_2})}
={n_{12}!m_{12}!(n_1+m_1)!(n_2+m_2)!\over n_1!n_2!m_1!m_2!(n_{12}+m_{12})!}.$$
Notice that these restricted Littlewood-Richardson numbers are all $\le 1$.

A second case of interest, are representations where $R$, $r_n$ and $r_m$ are totally symmetric. The corresponding single matrix correlators 
were conjectured to be dual to branes that have expanded in the AdS$_5$ of the AdS$_5\times$S$^5$ geometry. A beautiful test of this conjecture was
given in \cite{Correa:2006yu}. We again conjecture that the same interpretation holds even when the operator is built using
two complex Higgs fields. Again, this class of restricted characters are easy to compute because the dimension of the representation
and the dimension of the restriction are equal. Exactly as above, we can express the restricted Littlewood-Richardson numbers in terms of the 
Littlewood-Richardson numbers as
\begin{eqnarray} 
f^{(n_{12}+m_{12}),((n_1+n_2),(m_1+m_2))}_{(n_1+m_1),((n_1),(m_1))\, (n_2+m_2),((n_2),(m_2))}
&=&{n_{12}!m_{12}!(n_1+m_1)!(n_2+m_2)!\over n_1!n_2!m_1!m_2!(n_{12}+m_{12})!}f_{(n_1+m_1)\, (n_2+m_2)\, (n_{12}+m_{12})}
\nonumber\\
&=&{n_{12}!m_{12}!(n_1+m_1)!(n_2+m_2)!\over n_1!n_2!m_1!m_2!(n_{12}+m_{12})!},\nonumber
\end{eqnarray}
where we have used the fact that in this case the Littlewood-Richardson number is again 1.
Again, upon setting $m_1=m_2=0$ our restricted Littlewood-Richardson numbers reproduce the standard Littlewood-Richardson numbers.

Another case in which general results can be obtained, involves multiplying a polynomial in $Z$ ($\chi_{R_1}(Z)$ with $R_1$ an
irreducible representation of $S_n$) by a polynomial in $X$ ($\chi_{R_2}(X)$ with $R_2$ an irreducible representation of $S_m$).
The relevant restricted Littlewood-Richardson numbers are simply
\begin{eqnarray}
f^{R,(r_n,r_m)}_{R_1,(R_1,\cdot)\, R_2,(\cdot,R_2)}&=&{d_R\over (n+m)!d_{r_n}d_{r_m}}
\sum_{\sigma_1\in S_n}\sum_{\sigma_2\in S_m}\chi_{R_1}(\sigma_1)\chi_{R_2}(\sigma_2)
\chi_{R,(r_n,r_m)}(\sigma_1\circ\sigma_2)\nonumber\\
&=&{d_R\over (n+m)!d_{r_n}d_{r_m}}
\sum_{\sigma_1\in S_n}\chi_{R_1}(\sigma_1)\chi_{r_n}(\sigma_1)
\sum_{\sigma_2\in S_m}\chi_{R_2}(\sigma_2)\chi_{r_m}(\sigma_2)\nonumber\\
&=&{n!m!\over (n+m)!}{d_R\over d_{r_n}d_{r_m}}\delta_{R_1 r_n}\delta_{R_2 r_m}\, .
\nonumber
\end{eqnarray}
We have not explicitly indicated a multiplicity label for $r_n$ and $r_m$. This label is needed if there is more than
one way to subduce $(r_n,r_m)$ from $R$.

Finally, if we consider products of polynomials for which the total number of $X$s participating is small, we can construct the
explicit change of basis developed in the previous subsection. With this change of basis in hand, all restricted characters 
appearing in the formula for the restricted Littlewood-Richardson numbers can be computed using standard strand diagram
techniques.

\section{Graviton Interactions}

A description of perturbative quantum gravity phrased in terms of graviton degrees of freedom is expected to fail at high 
enough energy, when the gravitons become strongly interacting. A perturbative description of the high energy physics needs 
to be phrased in terms of new degrees of freedom that are weakly interacting at high energy. The exact computations of
\cite{Dhar:2005su}, performed in the half-BPS sector of IIB string theory on AdS$_5\times$S$^5$ show that the sphere giant
gravitons provide satisfactory degrees of freedom for a high energy description. The half-BPS dynamics is captured by the
dynamics of a single matrix. In this section, using the methods we have developed above, 
we will extend the results of \cite{Dhar:2005su} beyond the one matrix sector.
Our conclusions are consistent with those of \cite{Dhar:2005su}.

To estimate where the graviton description breaks down, we will estimate where the two point functions differ appreciably from
their planar limit. This difference indicates that non-planar corrections are no longer suppressed and this implies that
the large $N$ orthogonality of the trace basis fails.
The trace basis is dual, at low energies, to the supergravity Fock space. The fact that orthogonality fails tells us that there
are non-negligible interactions mixing the graviton Fock space states; these interactions will invalidate perturbation theory. If we build
traces that contain only $X$s or only $Z$s, we know that orthogonality fails when we have $O(\sqrt{N})$ fields in each 
trace\cite{Beisert:2002bb,Constable:2002hw}. Is this conclusion still valid when we consider traces containing both $X$s and $Z$s?
The computations of \cite{Beisert:2002bb,Constable:2002hw} used the fact that the combinatorics associated with the Wick contractions
can be computed using a zero dimensional complex matrix model. These techniques do not have an easy extension to the mixed traces
that interest us. A more powerful alternative technique (employing symmetric group theory methods) to compute the single trace 
correlators was developed in \cite{Corley:2002mj}. Using the logic of \cite{Corley:2002mj}, we are able to compute mixed trace 
correlators. The basic idea is that using the results of \cite{Bhattacharyya:2008rb}, we know how to compute mixed correlators in 
the restricted Schur basis. To get correlators of traces we simply need to change from the restricted Schur basis to the trace basis. 
This change of basis is easily accomplished using dual characters. Any single trace operator can be written as
$$\Tr (\sigma_c Z^{\otimes n}\otimes X^{\otimes m})=\sum_{R,(r_n,r_m)}
\chi^{R,(r_n,r_m)}(\sigma_c)\chi_{R,(r_n,r_m)}(Z,X),$$
where $\sigma_c$ is an $n+m$ cycle. For concreteness we will consider the trace $\Tr (Z^nX^m)$ which corresponds to
$\sigma_c=(1\,2\,3\,\cdots\, m+n-2\, m+n-1\, m+n)$, up to conjugation by an element of $S_n\times S_m$. The computation of the relevant
restricted characters is carried out in appendix D. We find that this character vanishes unless $r_n$ and $r_m$ are both hooks - Young
diagrams with at most a single column whose length is $>1$. Denote the number of rows in $r_n$ by $s_n+1$ and the number of rows in $r_m$ by $s_m+1$.
To subduce $(r_n,r_m)$, the Young diagram $R$ must have at most two column with lengths $>2$. We again imagine that boxes are removed 
from $R$ to compose $r_m$. The boxes that remain form $r_n$. The result of appendix D says
$$
\chi_{R,(r_n,r_m)}(\sigma_c)={(-1)^{s_n+s_m}\over mn}\left(\sum_i c_i -mN-{m(m-2s_m -1)\over 2}\right),
$$
where the sum on $i$ runs over all boxes removed from $R$ to produce $r_m$. Thus, the dual character is
\begin{eqnarray}
\chi^{R,(r_n,r_m)}(\sigma_c)
&=&{{\rm (hooks)}_{r_n}{\rm (hooks)}_{r_m}\over {\rm (hooks)}_{R}}
{(-1)^{s_n+s_m}\over mn}\left(\sum_i c_i -mN-{m(m-2s_m -1)\over 2}\right)\nonumber\\
&=& s_n! (n-s_n-1)!s_m! (m-s_m-1)!(-1)^{s_n+s_m}\times {\cal F},\nonumber
\end{eqnarray}
where the $R$ dependent factor ${\cal F}$ is given by
$$ {\cal F} = {\left(\sum_i c_i -mN-{m(m-2s_m -1)\over 2}\right)\over {\rm (hooks)}_{R} }.$$
Using the two point functions of \cite{Bhattacharyya:2008rb} it is now straight forward to obtain
$$ \langle \Tr (\sigma_c Z^{\otimes n}\otimes X^{\otimes m})^\dagger \Tr (\sigma_c Z^{\otimes n}\otimes X^{\otimes m})\rangle
=\sum_{R,(r_n,r_m)}{s_m! (m-s_m-1)! s_n! (n-s_n-1)!\over mn}$$
$$\times {\rm Dim}_R
\left(\sum_i c_i -mN-{m(m-2s_m -1)\over 2}\right)^2,$$
where ${\rm Dim}_R$ is the dimension of $SU(N)$ representation $R$ and the sum runs over all labels such that $r_n$ and $r_m$ are hooks.
Denote the number of boxes in the first row of $R$ by $r_1$ and the number of boxes in the second row of $R$ by $r_2$.
Denote the number of boxes in the first column of $R$ by $c_1$ and the number of boxes in the second column of $R$ by $c_2$. It is
straight forward to see that
$$ r_2=m+n-s_m-s_n-r_1+1,\qquad c_2=s_n+s_m-3-c_1,\qquad r_1+r_2+c_1+c_2=n+m+4 .$$
Summing over $R,(r_n,r_m)$ can be replaced by a sum over $s_n$, $s_m$, $r_1$ and $c_1$. The value of the sum is
$$ \langle \Tr (\sigma_c Z^{\otimes n}\otimes X^{\otimes m})^\dagger \Tr (\sigma_c Z^{\otimes n}\otimes X^{\otimes m})\rangle ={N^2\over (n+1)(m+1)(N^2-1)}\times$$
\begin{eqnarray}
&\, & 
\left({1\over (n+2)(m+2)}
\left[{(N+n+1)!\over N!}-{(N-1)!\over (N-n-2)!}\right]
\left[{(N+m+1)!\over N!}-{(N-1)!\over (N-m-2)!}\right]
\right.
\nonumber\\
&-&{1\over N(m+2)}
\left[{(N+n)!\over N!}-{(N-1)!\over (N-n-1)!}\right]
\left[{(N+m+1)!\over N!}-{(N-1)!\over (N-m-2)!}\right]
\nonumber\\
&-&{1\over N(n+2)}
\left[{(N+n+1)!\over N!}-{(N-1)!\over (N-n-2)!}\right]
\left[{(N+m)!\over N!}-{(N-1)!\over (N-m-1)!}\right]
\nonumber\\
&+&\left.
\left[{(N+n)!\over N!}-{(N-1)!\over (N-n-1)!}\right]
\left[{(N+m)!\over N!}-{(N-1)!\over (N-m-1)!}\right]\right)\, .
\nonumber
\end{eqnarray}
The above result is exact (i.e. to all orders in $N$) in the free field theory limit.
Following \cite{Dhar:2005su} we make us of the identity
$${(N+p_1)!\over (N-p_0-1)!}=N^{p_0+p_1+1}\exp\left[{1\over 2N}(p_1-p_0)(p_0+p_1+1)+O(N^{-2})\right],$$
valid when $p_0\ll N$ and $p_1\ll N$, to explore the behavior of our two point function.
We find an appreciable difference from the planar answer (which is $N^{n+m}$) when $n^2/ N$ or $m^2/ N$ (or both) are held fixed.
The leading behaviour of our two point correlation function is
$$ \langle \Tr (\sigma_c Z^{\otimes n}\otimes X^{\otimes m})^\dagger \Tr (\sigma_c Z^{\otimes n}\otimes X^{\otimes m})\rangle =
{4N^{n+m+2}\over nm}\sinh {n^2\over 2N}\sinh {m^2\over 2N}.$$
We have dropped terms of order ${m\over N}$ and ${n\over N}$.
No finite order calculation in $1/N$ will reproduce this result - in this range of values for $m,n$ perturbation theory breaks down.
This suggests that, even when more than one matrix species participates in the trace, the validity of perturbation theory in terms of
gravitons (that is single trace operators) breaks down when there are $O(\sqrt{N})$ matrices in the trace. The cautious reader
might object that we have not built operators that correspond to chiral primaries of the SYM and hence that our operators are
not dual to gravitons. This is indeed true. However, the effects we study arise because the large number of non-planar contractions
possible over power the usual $N^{-2}$ suppression of higher genus effects. Thus, this effect is really only sensitive to the number
of fields appearing in each trace and not the specific details of how these fields are distributed.

We will now compute the three point function for three restricted Schur polynomials. In what follows we will always employ the multi particle
normalization 
$$ \Gamma (1;2 | 1+2)=
{\langle \chi_{\{1\}}\chi_{\{2\}}\chi_{\{1+2\}}^\dagger \rangle
\over ||\chi_{\{1\}}||\, ||\chi_{\{2\}}||\, ||\chi_{\{1+2 \}}|| }$$
of \cite{Brown:2006zk}. Consider first the case that each restricted Schur has totally antisymmetric labels i.e. labels of the form 
$1^{n+m},(1^n,1^m)$. This corresponds to the interaction of three sphere giant gravitons.
In this case it is straight forward, using the results of section 2 and 3.3 and of \cite{Bhattacharyya:2008rb} to 
obtain
\begin{eqnarray} 
\Gamma (1;2 | 1+2)&=&
{\langle \chi_{1^{n_1+m_1},(1^{n_1},1^{m_1})}\chi_{1^{n_2+m_2},(1^{n_2},1^{m_2})}\chi^\dagger_{1^{n_{12}+m_{12}},(1^{n_{12}},1^{m_{12}})}\rangle
\over ||\chi_{1^{n_1+m_1},(1^{n_1},1^{m_1})}||\, ||\chi_{1^{n_2+m_2},(1^{n_2},1^{m_2})}||\,
||\chi_{1^{n_{12}+m_{12}},(1^{n_{12}},1^{m_{12}})}||}\nonumber\\
&=&\sqrt{(N-n_1-m_1)!(N-n_2-m_2)!\over N! (N-n_{12}-m_{12})!}\sqrt{n_{12}!m_{12}!(n_1+m_1)!(n_2+m_2)!\over n_1!n_2!m_1!m_2!(n_{12}+m_{12})!}\, .
\nonumber
\end{eqnarray}
We have written this result as a product of two square root factors. The first factor on the last line has the same form as the one matrix result.
If $m_1=m_2=0$, this factor is identically equal to the correlation function computed in the one matrix case. The growth of this term for different
values of $n_1+m_1$ and $n_2+m_2$ has been considered in detail in \cite{Corley:2001zk,Dhar:2005su}. For both $n_1+m_1$ and $n_2+m_2$ fixed as $N\to\infty$
we find that\cite{Corley:2001zk,Dhar:2005su} $\Gamma (1;2 | 1+2)\sim O(1)$, so that these restricted Schur polynomials do not provide weakly
coupled degrees of freedom for long wavelength (low energy) modes. For ${n_1+m_1\over N}$ and ${n_2+m_2\over N}$ fixed and small in the limit $N\to\infty$
we find that\cite{Corley:2001zk,Dhar:2005su} $\Gamma (1;2 | 1+2)\sim e^{-\alpha N}$ with $\alpha$ positive and $O(1)$: these restricted Schur polynomials 
provide weakly coupled degrees of freedom for high energy modes. The second factor is always $\le 1$. To see this, consider the binomial expansion of 
$$ (1+x)^m = \sum^m_{k=0}\left(^m_k\right)x^k,\qquad {\rm where}\qquad \left(^m_k\right)={m!\over k!(m-k)!}.$$
By comparing the coefficient of $x^{r+s}$ coming from the expansion of $(1+x)^m$ times the expansion of $(1+x)^n$ to the coefficient of
$x^{r+s}$ coming from the expansion of $(1+x)^{m+n}$, we learn that
$$ \left(^m_r\right)\left(^n_s\right)\quad + \quad {\rm non\, negative\, integers}\quad = \left(^{m+n}_{r+s}\right).$$
Thus
$$ {\left(^m_r\right)\left(^n_s\right)\over\left(^{m+n}_{r+s}\right)} \le 1,$$
which proves that the second factor is $\le 1$. Notice that when $m_1=m_2=0$, the second factor
is identically equal to 1 so that our result correctly reduces to the one matrix result of \cite{Corley:2001zk}. In summary, in the multimatrix sector 
we find the giant graviton degrees of freedom are weakly coupled at high energy, consistent with the conclusions of \cite{Dhar:2005su}.

Finally, it is equally easy to compute the correlation function in the case that all three giants interacting are in completely symmetric representations.
This corresponds to the interaction of three AdS giant gravitons. The result is
\begin{eqnarray} 
\Gamma (1;2 | 1+2)&=&
{\langle \chi_{(n_1+m_1),((n_1),(m_1))}\chi_{(n_2+m_2),((n_2),(m_2))}\chi^\dagger_{(n_{12}+m_{12}),((n_{12}),(m_{12}))}\rangle
\over ||\chi_{(n_1+m_1),((n_1),(m_1))}||\, ||\chi_{(n_2+m_2),((n_2),(m_2))}||\,
||\chi_{(n_{12}+m_{12}),((n_{12}),(m_{12}))}||}\nonumber\\
&=&\sqrt{(N+n_{12}+m_{12}-1)!(N-1)!\over (N+m_1+n_1-1)! (N+n_{2}+m_{2}-1)!}\sqrt{n_{12}!m_{12}!(n_1+m_1)!(n_2+m_2)!\over n_1!n_2!m_1!m_2!(n_{12}+m_{12})!}\, .
\nonumber
\end{eqnarray}
It is again easy to verify that if we set $m_1=0=m_2$, we recover the one matrix result of \cite{Corley:2001zk}. For ${n_1+m_1\over N}$ and ${n_2+m_2\over N}$ 
fixed and small in the limit $N\to\infty$, we find $\Gamma (1;2 | 1+2)\sim e^{\alpha N}$ with $\alpha$ positive and $O(1)$ suggesting that these restricted Schur 
polynomials do not provide weakly coupled degrees of freedom for high energy modes.

\section{Interpretation of Restricted Schur Polynomials}

The restricted Schur polynomial $\chi_{R,(r_n,r_m)}(Z,X)$ has three labels $R$, $r_n$ and $r_m$. The label $r_n$ is naturally associated to the $Z$s and the
label $r_m$ to the $X$s. It seems natural to think that the composite operator $\chi_{R,(r_n,r_m)}(Z,X)$ is constructed from the half
BPS ``partons'' $\chi_{r_n}(Z)$ and $\chi_{r_m}(X)$. In this section we will see if we can gather some evidence that $\chi_{R,(r_n,r_m)}(Z,X)$
indeed has a partonic structure.

Before considering the parton structure of the restricted Schur polynomials, we recall a related relevant problem: the
parton structure of hadrons\cite{Peskin}. At high energy, above about 10 GeV, proton-proton scattering produces a large number of pions, with momenta
mainly collinear with the collision axis. The probability of producing a pion with a large component of momentum transverse to the collision axis is
exponentially suppressed with the value of the transverse momentum. A picture of the proton as a loosely bound assemblage of partons was consistent
with the observed data. It is at high energies, where the partons become weakly interacting (thanks to asymptotic freedom) that they are visible
in the experimental results. At low energies the constituents are strongly interacting and the parton structure of the proton is not visible.

{\vskip 0.25cm}

{\sl Is there an analogous limit in which we expect the constituents of the restricted Schur polynomial are weakly interacting? Do we see evidence for a 
partonic structure in this limit? }

{\vskip 0.25cm}

There are two distinct ways in which we could obtain a weakly interacting system.
The parameter which controls the interactions among our proposed constituents is ${1\over N}$. We could thus imagine taking $N\to\infty$ and
systematically expanding our (exact) correlator results, keeping only the leading order. Alternatively, we could ask if there are situations
where we expect the constituents are naturally weakly interacting. We will follow this second approach. Interactions between membranes are
mediated by the open strings ending on the membrane's worldvolume; in general the string and the membrane can exchange momentum, that is, the string
can exert a force on the membrane. For the special case of a nearly maximal sphere giant\footnote{To be more precise, we consider a giant graviton 
which carries momentum $p$ with $N-p$ a number of $O(1)$.}, this interaction is highly suppressed and the open strings attached to a maximal giant 
do not exert a force on it\cite{Berenstein:2006qk}. Thus, to get weakly interacting partons, 
we consider a bound state built from two partons, one of which is a boundstate
of nearly maximal sphere giants. In what follows, we take $r_n$ to be a Young diagram with $c$ columns and $p$ rows with $c$ a number of $O(1)$ and
$N-p$ a number of $O(1)$. From the results of section 3.3 we know that
$$\chi_{r_n}(Z)\chi_{r_m}(X)={n!m!\over (n+m)!d_{r_n}d_{r_m}}\sum_{R}\sum_{i} \, d_R\chi_{R,(r^{(i)}_n,r^{(i)}_m)}(Z,X),$$
where the sum is over all representations $R$ that can subduce $(r_n,r_m)$ and $(i)$ is a multiplicty label distinguishing the different
$(r_n,r_m)$ representations that can be subduced. If we normalize the above operators so that they have a unit two point function we obtain
\begin{equation}
\hat{\chi}_{r_n}(Z)\hat{\chi}_{r_m}(X)=\sum_{R}\sum_{i} \, \sqrt{{\rm Dim}_R\over {\rm Dim}_{r_n}{\rm Dim}_{r_m}}
\hat{\chi}_{R,(r^{(i)}_n,r^{(i)}_m)}(Z,X),
\label{atoms}
\end{equation}
where a hat denotes operators with unit two point function and ${\rm Dim}_R$ denotes the dimension of the $SU(N)$ irreducible representation
labeled by $R$. It is simple to check that there is a single Young diagram $R$ that dominates
the sum on the right hand side; it has the form displayed in figure 1 below. 
\myfig{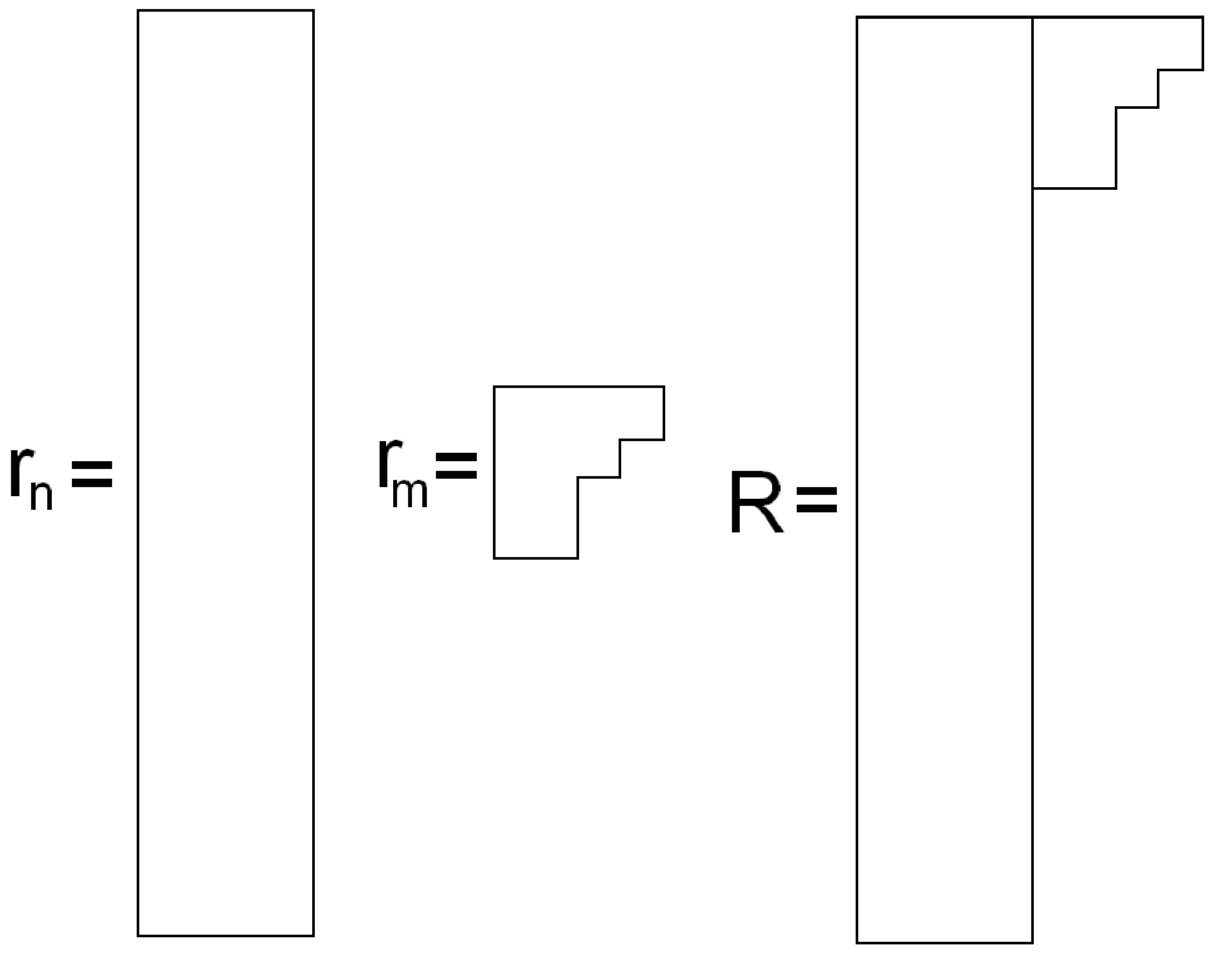}{10.0}{The Young diagram which dominates (\ref{atoms}) is obtained by stacking $r_m$ next to $r_n$ as shown.}
Indeed, any boxes stacked below $r_n$ must be antisymmetrized
with the other indices in the same column of the Young diagram; since the number of boxes already in the column is $p$, the index corresponding
to the boxes stacked below $r_n$ can only take $N-p = O(1)$ values. In addition, one divides by a big symmetry factor (it is $O(N)$). In contrast to this, 
the indices corresponding to boxes stacked next to $r_n$ are symmetrized with boxes appearing in the same row. These can take $O(N)$ values.
Since the number of boxes in the row is $O(1)$ one divides by an extra symmetry factor of $O(1)$. There is a single way to subduce this dominant
$R$ so that there is no need for the $i$ index. It is now clear that
the leading correction to this term is of order ${1\over N}$. Lets now consider the interaction of $\hat{\chi}_{R,(r_{n_1},r_{m_1})}(Z,X)$
and $\hat{\chi}_{S,(s_{n_2},s_{m_2})}(Z,X)$, where both $r_{n_1}$ and $s_{n_2}$ are boundstates of nearly maximal sphere giants. This is
the analog of scattering two protons. Using (\ref{atoms}) it is now straight forward to see that
$$ \chi_{R,(r_{n_1},r_{m_1})}(Z,X)\chi_{S,(s_{n_2},s_{m_2})}(Z,X)
={({\rm hooks})_R({\rm hooks})_S \over ({\rm hooks})_{r_{n_1}}({\rm hooks})_{r_{m_1}}({\rm hooks})_{s_{n_2}}({\rm hooks})_{s_{m_2}}}\times$$
$$\times\sum_{t_{n_{12}}}\sum_{t_{m_{12}}}\sum_{T}\sum_i {n_{12}! m_{12}! d_T\over (n_{12}+m_{12})! d_{t_{n_{12}}}d_{t_{m_{12}}}}
f_{r_{n_1}s_{n_2}t_{n_{12}}}f_{r_{m_1}s_{m_2}t_{m_{12}}}
\chi_{T,(t_{n_{12}}^{(i)},t_{m_{12}}^{(i)})}(Z,X)\, .$$
Even in the large $N$ limit, the right hand side is a sum over a number of terms - many possible states can be formed as a result of the
interaction of our two composites. Note however that the detailed structure of this interaction, in particular which
representations $t_{n_{12}}$ and $t_{m_{12}}$ appear, is determined by the Littlewood-Richardson
coefficients $f_{r_{n_1}s_{n_2}t_{n_1+n_2}}$ and $f_{r_{m_1}s_{m_2}t_{m_1+m_2}}$. These Littlewood-Richardson coefficients give the
detailed form of the interaction between the ${1\over 2}$ BPS partons
$$ \chi_{r_{n_1}}(Z) \chi_{s_{n_2}}(Z)=\sum_{t_{n_1+n_2}}f_{r_{n_1}s_{n_2}t_{n_1+n_2}}\chi_{t_{n_1+n_2}}(Z),$$
$$ \chi_{r_{m_1}}(X) \chi_{s_{m_2}}(X)=\sum_{t_{m_1+m_2}}f_{r_{m_1}s_{m_2}t_{m_1+m_2}}\chi_{t_{m_1+m_2}}(X).$$
Thus, the interaction of the composites is as a result of interactions between the partons. The picture that has emerged is very similar
to the parton structure of hadrons described above.

\section{Discussion}

In this article we have demonstrated that the restricted Schur polynomials satisfy a simple product rule, generalizing the rule known for
Schur polynomials. Using this rule it is now possible to perform computations of higher point correlation functions of restricted Schur 
polynomials. We have obtained explicit formulas for three point functions of a large class of restricted Schur polynomials which are built
using antisymmetric or symmetric representations. Using these results we have been able to argue that restricted Schur polynomials built
using the antisymmetric representations provide a suitable set of degrees of freedom for the description of perturbative quantum gravity,
in agreement with the conclusions of \cite{Dhar:2005su} and extending the explicit computations performed there.

A key building block appearing in the restricted Schur polynomial is the restricted character.
The restricted character is a generalization of the usual character, which plays a central role in group theory. In this paper there is a 
development of the restricted character theory parallel to the character theory of finite groups. The notions of a conjugacy 
class and of a dual character, have been generalized to the notions of restricted conjugacy class and dual restricted characters. These
generalizations lead to a set of orthogonality relations satisfied by the restricted characters. In this way we have ultimately obtained 
the generalization of the Littlewood-Richardson coefficient.

By studying the interaction of two restricted Schur polynomials we have suggested a physical interpretation for the labels of the restricted
Schur polynomial: the composite operator $\chi_{R,(r_n,r_m)}(Z,X)$ 
is constructed from the half BPS ``partons'' $\chi_{r_n}(Z)$ and $\chi_{r_m}(X)$. Specifically, we have identified a composite operator in 
which the constituent partons are weakly interacting. The interaction of two such composites is largely determined by the interactions of
the partons. This is analogous to the shower of pions produced when two hadrons interact; the pion shower can be attributed to the parton-parton
interactions between pairs of partons belonging to different hadrons. In cases when the partons are strongly 
interacting (for example, at low energy), they are not visible and a partonic description of hadrons is not useful. 
In a similar way, we expect that when the constituent half BPS partons are strongly interacting, this might not be a useful interpretation 
of the restricted Schur polynomials. The picture of a composite comprised of half BPS partons seems to be similar to the proposal of 
\cite{Balasubramanian:2006gi} that every supersymmetric four dimensional black hole of finite area can be split up into microstates made 
of primitive half-BPS ``atoms''. The idea that each half BPS state should be treated as an independent parton matches nicely with the picture 
that has emerged for half-BPS states in AdS$_5$\cite{Lin:2004nb,Balasubramanian:2005kk}. Finally, it is not yet known how to build restricted Schur polynomials
that correspond to ${1\over 4}$ or ${1\over 8}$ BPS states. In \cite{Mandal:2006tk} ${1\over 8}$ and ${1\over 4}$ BPS states were obtained 
by putting together ${1\over 2}$ BPS dual giants; making contact with that work may indicate which restricted Schur polynomials 
correspond to ${1\over 4}$ or ${1\over 8}$ BPS states.

Although we have focused on the case of two matrices, the extension to more matrices is straight forward as described in Appendix C. 
It would be interesting to work out some explicit examples with more than two matrices. Further, even for two matrices, it would be
nice to make the formalism developed here more efficient. Indeed, in computing the values of two cycles with one index in the $S_n$
subgroup and one index in the $S_m$ subgroup, we have built a change of basis. This allows us not only to compute the value of the trace,
but the value of any matrix element. This is much more information than we actually need; surely more efficient methods can be found.

The Littlewood-Richardson numbers have many interpretations: as coefficients in the decomposition of tensor products into
irreducible $GL_n$ modules, as coefficients in the decomposition of skew Specht modules into irreducibles, as coefficients
in the decomposition of $S_n$ representations induced from Young subgroups and as intersection numbers in the Schubert
calculus on a Grassmannian\cite{Stanley}. How much of this generalizes for restricted Littlewood-Richardson numbers? Finally,
our work contains some general directions for the study of multi-matrix models. The eigenvalue based techniques that were
so useful for the description of one matrix models do not seem to have an easy generalization for multi matrices. For one matrix
models, the eigenvalues provide a set of $O(N)$ variables; since fluctuations about the large $N$ configuration are $O(N^{-2})$,
these provide a suitable set of variables for the implementation of a saddle point approximation.
In general, it would be incorrect to assume that the matrices of a multimatrix model can be simultaneously diagonalized\footnote{An
exception to this is the sector of the theory comprising of the chiral ring at strong coupling - see \cite{Berenstein}.}. Thus, at best 
it seems that one needs to keep
the eigenvalues of each matrix, plus unitary matrices which encode how one goes between the bases in which a particular matrix is
diagonal. This gives a total of $O(N^2)$ variables, so that this does not provide a promising starting point for a saddle point
analysis. In contrast to this, 
using the technology of \cite{de Mello Koch:2007uu,Brown:2007xh,Kimura:2007wy,Bhattacharyya:2008rb}, we can give a rather 
detailed description of the free multi-matrix models. One loop results for the super Yang-Mills F term also have a description in this 
framework\cite{de Mello Koch:2007uv,Bekker:2007ea,Brown:2008rs}. It would be interesting to develop these methods further
by developing efficient techniques for extracting large $N$ results and for managing more general potentials.

{\vskip 1.0cm}

\noindent
{\it Acknowledgements:} We would like to thank Tom Brown, Jeff Murugan, Sanjaye Ramgoolam and Joao Rodrigues for enjoyable, 
helpful discussions. This work is based upon research supported by the South African Research Chairs Initiative 
of the Department of Science and Technology and National Research Foundation. Any opinion, findings and conclusions 
or recommendations expressed in this material are those of the authors and therefore the NRF and DST do not accept 
any liability with regard thereto. This work is also supported by NRF grant number Gun 2047219.

\appendix

\section{Identities and notation}

The polynomials we construct are built out of two matrices, $X$ and $Z$. We will typically use an $n$ to denote the number of $Z$s
in the polynomial and an $m$ to denote the number of $X$s in the polynomial. The polynomial is built using an irreducible representation
of the symmetric group $S_{n+m}$ that permutes the indices of the $Z$ and $X$ fields amongst each other. We will denote the Young
diagrams labeling representations of the symmetric group acting on both $X$s and $Z$s using capital letters. The indices of the
matrix representing an element of this group will be denoted by capital Greek letters. Thus, the elements of the matrix representing 
$\sigma$ in the $S_{n+m}$ irreducible representation $R$ will be denoted by $\big[\Gamma_R (\sigma)\big]_{\Psi\Phi}$.
The irreducible representations of the $S_n$ subgroup that acts only on the indices of the $Z$ fields will be denoted by a small
letter with a subscript $n$; matrix elements of this representation will be indexed using letters from the start of the alphabet. 
Thus, the elements of the matrix representing $\sigma$ in the $S_{n}$ irreducible representation $r_n$ will be denoted by 
$\big[\Gamma_{r_n} (\sigma)\big]_{ab}$.
The irreducible representations of the $S_m$ subgroup that acts only on the indices of the $X$ fields will be denoted by a small
letter with a subscript $m$; matrix elements of this representation will be indexed using letters from the middle of the alphabet. 
Thus, the elements of the matrix representing $\sigma$ in the $S_{m}$ irreducible representation $r_m$ will be denoted by 
$\big[\Gamma_{r_m} (\sigma)\big]_{ij}$. Finally, irreducible representations of the $S_n\times S_m$ subgroup that permute the
indices of the $X$s and permute the indices of the $Z$s is denoted by $(r_n,r_m)$.

The delta function $\delta(\sigma )$ is $1$ if $\sigma$ is the identity and zero otherwise. In this article the $\delta (\cdot )$ is
usually summed with a summand that is a class function. In addition, we usually have a delta function defined on the classes, i.e. we
have $\delta (\big[\sigma\big]_r\big[\tau\big]_r )$ instead of $\delta (\sigma\tau )$. The relation between these two is easily
established: Let $F(\big[\sigma ]_r)$ be any class function. By comparing 
$$\sum_{\sigma\in S_n} F(\big[\sigma ]_r)\delta (\big[\sigma\big]_r\big[\tau\big]_r )\qquad {\rm and}\qquad
  \sum_{\sigma\in S_n} F(\big[\sigma ]_r)\delta (\sigma\tau ) $$
we find that, when summing the delta function multiplied by any class function, over the whole group, we can freely replace
$$ {\delta (\big[\sigma\big]_r\big[\tau\big]_r )\over n^\sigma_{R, (r_n,r_m)}}\leftrightarrow\delta (\sigma\tau). $$

\section{A Formula for the restricted Littlewood-Richardson numbers}

In this appendix we will employ a bra/ket notation. 
In what follows, $R$ is an irreducible representation of $S_{n+m}$ and $(r_{n},r_{m})$ is an
irreducible representation of $S_n\times S_m$. We label the irreducible representations of $S_n\times S_m$
by a Young diagram $r_{n}$ with $n$ boxes (which labels an irreducible representation of $S_n$) and
a Young diagram $r_{m}$ with $m$ boxes (which labels an irreducible representation of $S_m$). For 
states belonging to the carrier space of $R$ we write $|R,\Gamma\rangle$. For states belonging to the carrier
space of $(r_{n},r_{m})$ we write labels acted on by the $S_n$ and $S_m$ subgroups separately
$|r_{n},a;r_{m},i\rangle $. For example, we write
$$\big[\Gamma_R (\alpha )\big]_{\Lambda\Psi}=\langle R ,\Lambda|\alpha |R ,\Psi\rangle .$$
For a trace over an ``on the diagonal block'' (see \cite{Balasubramanian:2004nb},\cite{de Mello Koch:2007uu} for an
explanation of this terminology) we write
$$\Tr_{(r_{n},r_{m})}(\Gamma_R(\sigma))=\sum_{i,a}\sum_{\Lambda\Theta}\langle r_{n},a;r_{m},i|R,\Lambda\rangle
\big[\Gamma_{R}(\sigma)\big]_{\Lambda\Theta}\langle R,\Theta |r_{n},a;r_{m},i\rangle .$$
For an ``off the diagonal block'' we write
$$\Tr_{(r_{n},r_{m}),(s_{n},s_{m})}(\Gamma_R(\sigma))=\sum_{i,a}\sum_{\Lambda\Theta}\langle r_{n},a;r_{m},i|R,\Lambda\rangle
\big[\Gamma_{R}(\sigma)\big]_{\Lambda\Theta}\langle R,\Theta |s_{n},a;s_{m},i\rangle .$$
For the labels of the off the diagonal block, we need $(r_{n},r_{m})$ to have the same shape as $(s_{n},s_{m})$. This means that
$r_{n}$ and $s_{n}$ as well as $s_{m}$ and $r_{m}$ have the same shape.

For $\alpha_1\in S_n$ and $\alpha_2\in S_m$ we have
\begin{eqnarray}
\big[\Gamma_R (\alpha_1\circ\alpha_2)\big]_{\Psi\Theta}&=&\langle R,\Psi|\alpha_1\circ\alpha_2 |R,\Theta\rangle\nonumber\\
&=&\sum_{r_{n},r_{m},a,i}\sum_{t_n,t_m,b,j}
\langle R,\Psi |r_{n},a;r_{m},i\rangle\langle r_{n},a;r_{m},i|\alpha_1\circ\alpha_2 
|t_n,b;t_m ,j\rangle\nonumber\\
& &\qquad\times\langle t_n,b;t_m ,j|R,\Theta\rangle \nonumber\\
&=&\sum_{r_{n},r_{m},a,i}\sum_{b,j}
\langle R,\Psi |r_{n},a;r_{m},i\rangle \big[\Gamma_{r_{n}}(\alpha_1)\big]_{ab}
\big[\Gamma_{r_{m}}(\alpha_2)\big]_{ij}\nonumber\\
& &\qquad\times\langle r_{n},b;r_{m},j|R,\Theta\rangle\, .\nonumber
\end{eqnarray}
To get to the last line we have used the fact that if $\alpha_1\in S_n$ and $\alpha_2\in S_m$ then
$\langle r_{n},a;r_{m},i|\alpha_1\circ\alpha_2 |t_n,b;t_m,j\rangle \propto$
$\delta_{r_{n}t_n}\delta_{r_{m}t_m}$. Use this identity to compute
\begin{eqnarray}
& &\sum_{\alpha_1\in S_n}\sum_{\alpha_2\in S_m}\big[\Gamma_R(\alpha_1\circ\alpha_2)\big]_{\Psi\Theta}
\big[\Gamma_{r_{n}}(\alpha_1^{-1})\big]_{ab}\big[\Gamma_{r_{m}}(\alpha_2^{-1})\big]_{ij}\nonumber\\
&=&\sum_{\alpha_1\, ,\alpha_2}\sum_{t_n t_m}\sum_{c\,d\,k\,l}
\langle R,\Psi |t_n ,c;t_m ,k\rangle
\big[\Gamma_{t_n}(\alpha_1)\big]_{cd}\big[\Gamma_{t_m}(\alpha_2)\big]_{kl}
\langle t_n,d;t_m ,l |R,\Theta\rangle
\big[\Gamma_{r_{n}}(\alpha_1^{-1})\big]_{ab}\big[\Gamma_{r_{m}}(\alpha_2^{-1})\big]_{ij}\nonumber\\
&=&{n!m!\over d_{r_{n}}d_{r_{m}}}\sum_i
\langle R,\Psi |r_{n}^{(i)},b;r_{m}^{(i)},j\rangle
\langle r_{n}^{(i)},a;r_{m}^{(i)},i |R,\Theta\rangle\, .
\nonumber
\end{eqnarray}
To obtain this result we have used the fundamental orthogonality relation
$$ \sum_{\alpha\in {\cal G}}\big[\Gamma_R(\alpha)\big]_{ab}\big[\Gamma_S(\alpha^{-1})\big]_{cd}={g\over d_R}\delta_{RS}
\delta_{ad}\delta_{bc},$$
with $d_R$ the dimension of irreducible representation $R$ and $g$ the order of ${\cal G}$. A given $S_n\times S_m$ irreducible
representation may be subduced more than once by an $S_{n+m}$ irreducible representation $R$. When applying this relation
to the sums over $\alpha_1$ and $\alpha_2$, we will get a contribution from all representations whose $S_n$ and $S_m$ labels
match. The index $i$ in $(r_{n}^{(i)},r_{m}^{(i)})$ runs over the complete set of identical $S_n\times S_m$ irreducible
representations. This last identity will be used to argue that the dual character is
$$\chi^{R,R_\gamma}(\tau )=\chi^{R,(r_{n}^{(i)},r_{m}^{(i)})(r_{n}^{(j)},r_{m}^{(j)})}(\tau )=
{d_R n! m!\over d_{r_{n}}d_{r_{m}}(n+m)! }
\chi_{R,(r_{n}^{(j)},r_{m}^{(j)})(r_{n}^{(i)},r_{m}^{(i)})}(\tau ).$$
To verify this, we compute
\begin{eqnarray}
& &\sum_{R,R_\gamma}\chi^{R,R_\gamma}(\tau )\chi_{R,R_\gamma}(\sigma )\nonumber\\
&=&\sum_R\sum_{r_{n}^{(i)},r_{m}^{(i)}}\sum_{r_{n}^{(j)},r_{m}^{(j)}}
\chi^{R,(r_{n}^{(i)},r_{m}^{(i)})(r_{n}^{(j)},r_{m}^{(j)})}(\tau )
\chi_{R,(r_{n}^{(i)},r_{m}^{(i)})(r_{n}^{(j)},r_{m}^{(j)})}(\sigma )\nonumber\\
&=&\sum_R\sum_{r_{n}^{(i)},r_{m}^{(i)}}\sum_{r_{n}^{(j)},r_{m}^{(j)}}
\sum_{\Theta ,\Psi ,\Gamma ,\Sigma}\sum_{a,b,i,j}\nonumber\\
& &\qquad\Gamma_R(\sigma)_{\Theta\Psi}\Gamma_R(\tau)_{\Gamma\Sigma}
\langle r_{n}^{(i)},a;r_{m}^{(i)},i|R,\Theta\rangle
\langle R,\Psi | r_{n}^{(j)},a;r_{m}^{(j)},i\rangle\nonumber\\
& &\qquad\times
\langle r_{n}^{(j)},b;r_{m}^{(j)},j|R,\Gamma\rangle
\langle R,\Sigma | r_{n}^{(i)},b;r_{m}^{(i)},j\rangle
{d_R n! m!\over d_{r_{n}}d_{r_{m}}(n+m)! }\nonumber\\
&=&\sum_R\sum_{r_{n},r_{m}}\sum_{\Theta ,\Psi ,\Gamma ,\Sigma}\sum_{a,b,i,j}
\Gamma_R(\sigma)_{\Theta\Psi}\Gamma_R(\tau)_{\Gamma\Sigma}
{d_R n! m!\over d_{r_{n}}d_{r_{m}}(n+m)!}
\left({d_{r_{n}}d_{r_{m}}\over n!m!}\right)^2\nonumber\\
& &\qquad\times\sum_{\alpha_1,\alpha_2}\big[\Gamma_R (\alpha_1\circ\alpha_2)\big]_{\Sigma\Theta}
\big[\Gamma_{r_{n}} (\alpha_1^{-1})\big]_{ba}\big[\Gamma_{r_{m}}(\alpha_2^{-1})\big]_{ji}
\nonumber\\
& &\qquad\times\sum_{\beta_1,\beta_2}\big[\Gamma_R (\beta_1\circ \beta_2)\big]_{\Psi\Gamma}
\big[\Gamma_{r_{n}} (\beta_1^{-1})\big]_{ab}\big[\Gamma_{r_{m}}(\beta_2^{-1})\big]_{ij}
\nonumber\\
&=&\sum_R \sum_{\alpha_1,\alpha_2}
\Gamma_R(\sigma)_{\Theta\Psi}\Gamma_R(\tau)_{\Gamma\Sigma}
{d_R\over (n+m)!}
\big[\Gamma_R (\alpha_1\circ\alpha_2)\big]_{\Sigma\Theta}
\big[\Gamma_R (\alpha_1^{-1}\circ\alpha_2^{-1})\big]_{\Psi\Gamma}
\nonumber\\
&=&\sum_R\sum_{\alpha_1,\alpha_2}
\chi_R(\sigma(\alpha_1^{-1}\circ\alpha_2^{-1})\tau(\alpha_1\circ\alpha_2))
{d_R \over (n+m)!}
\nonumber\\
&=& n!m!\delta (\big[\sigma\big]_r\big[\tau\big]_r),\nonumber
\end{eqnarray}
which demonstrates the result. In the above proof we have made use of the formula
$$\sum_R {d_R\over g}\chi_R(\sigma)=\delta(\sigma). $$
Clearly then,
$$ f_{R,R_\alpha\, S,S_\beta}^{T,(t_{n}^{(i)},t_{m}^{(i)})(t_{n}^{(j)},t_{m}^{(j)})}
= {1\over n_1! n_2! m_1! m_2!}\sum_{\sigma_1\in S_{n_1+m_1}}\sum_{\sigma_2\in S_{n_2+m_2}}$$
$${d_T (n_1+n_2)! (m_1+m_2)!\over d_{t_{n}}d_{t_{m}}(n_1+n_2+m_1+m_2)!}
\chi_{R,R_\alpha}(\sigma_1)\chi_{S,S_\beta}(\sigma_2)
\chi_{T,(t_{n}^{(j)},t_{m}^{(j)})(t_{n}^{(i)},t_m^{(i)})}(\sigma_1\circ\sigma_2).$$

\section{More than 2 Matrices}

Consider a matrix model with $d$ species of matrices.
$R$ is an irreducible representation of $S_{n_1+n_2+\cdots +n_d}$ and $(r_{n_1},r_{n_2},\cdots,r_{n_d})$ is an
irreducible representation of $S_{n_1}\times S_{n_2}\times\cdots\times S_{n_d}$. The Young diagram $r_{n_i}$ has $n_i$ 
boxes; it labels an irreducible representation of $S_{n_i}$. It is straight forward to show that

\begin{eqnarray}
& &\prod_{i=1}^d\sum_{\alpha_i\in S_{n_i}}
\big[\Gamma_R(\alpha_1\circ\alpha_2\circ\cdots\circ\alpha_d)\big]_{\Psi\Theta}
\big[\Gamma_{r_{{n_1}}}(\alpha_1^{-1})\big]_{a_1 b_1}\cdots\big[\Gamma_{r_{n_d}}(\alpha_d^{-1})\big]_{a_d b_d}\nonumber\\
&=&\prod_{i=1}^d{n_i!\over d_{r_{n_i}}}\sum_j
\langle R,\Psi |r_{n_1}^{(j)},b_1;r_{n_2}^{(j)},b_2;\,\cdots\,r_{n_d}^{(j)},b_d\rangle
\langle r_{n_1}^{(j)},a_1 ;r_{n_2}^{(j)}a_2;\,\cdots\,; r_{n_d}^{(j)}a_d|R,\Theta\rangle\, .
\nonumber
\end{eqnarray}
This last identity can again be used to argue that the dual character is
$$\chi^{R,(r_{n_1}^{(i)},r_{n_2}^{(i)},\cdots ,r_{n_d}^{(i)})(r_{n_1}^{(j)},r_{n_2}^{(j)},\cdots ,r_{n_d}^{(j)})}(\tau )=
\prod_{k=1}^d{n_k !\over d_{r_k}}{d_R\over (n_1+n_2+\cdots +n_k)! }
\chi_{R,(r_{n_1}^{(j)},r_{n_2}^{(j)},\cdots ,r_{n_d}^{(j)})(r_{n_1}^{(i)},r_{n_2}^{(i)},\cdots ,r_{n_d}^{(i)})}(\tau ).$$
Clearly then, ($R$ is an irreducible representation of $S_{n_1+n_2+\cdots+n_d}$; $R_\alpha$ is an irreducible representation
of $S_{n_1}\times S_{n_2}\times\cdots\times S_{n_d}$; $S$ is an irreducible representation of $S_{m_1+m_2+\cdots+m_d}$; 
$S_\beta$ is an irreducible representation of $S_{m_1}\times S_{m_2}\times\cdots\times S_{m_d}$; $T$ is an irreducible 
representation of $S_{n_1+n_2+\cdots+n_d+m_1+m_2+\cdots+m_d}$; $(r_{n_1}^{(i)},r_{n_2}^{(i)},\cdots ,r_{n_d}^{(i)})$ and
$(r_{n_1}^{(j)},r_{n_2}^{(j)},\cdots ,r_{n_d}^{(j)})$ have the same shape and are both irreducible representations
of $S_{n_1+m_1}\times S_{n_2+m_2}\times\cdots\times S_{n_d+m_d}$)
$$ f_{R,R_\alpha\, S,S_\beta}^{T,(r_{n_1}^{(i)},r_{n_2}^{(i)},\cdots ,r_{n_d}^{(i)})(r_{n_1}^{(j)},r_{n_2}^{(j)},\cdots ,r_{n_d}^{(j)})}
= \prod_{i=1}^d {1\over n_i! m_i!}\sum_{\sigma_1\in S_{n_1+n_2+\cdots+n_d}}\sum_{\sigma_2\in S_{m_1+m_2+\cdots+m_d}}$$
$$\prod_{k=1}^d{(n_k+m_k) !\over d_{r_k}}{d_R\over (n_1+n_2+\cdots +n_k+m_1+m_2+\cdots +m_k)! }\times$$
$$\times
\chi_{R,R_\alpha}(\sigma_1)\chi_{S,S_\beta}(\sigma_2)
\chi_{T,(r_{n_1}^{(j)},r_{n_2}^{(j)},\cdots ,r_{n_d}^{(j)})(r_{n_1}^{(i)},r_{n_2}^{(i)},\cdots ,r_{n_d}^{(i)})}(\tau ).$$

\section{Restricted Character Computations}

In this appendix we will consider the computation of the restricted character, for arbitrary representations, on the $m+n$ cycle
$\sigma_c=(1,2,3,\cdots,m+n)$. This character is needed to evaluate the two point functions used in section 4. Representations are 
labeled by a Young diagram. To specify a Young diagram, we will list the number of boxes in each row. In this appendix, the Young
diagrams, called ``hooks'' in \cite{Corley:2002mj} will play an important role. Listing the number of boxes in each row,
the hook diagrams are always of the form $(n+m-s,1,1,\cdots,1)$. 
Recall that the symmetric group character $\chi_R(\sigma_c)$ is $(-1)^s$ if $R$ is a hook and zero otherwise.
Let us verify this formula, using strand diagrams, for the hook ${\tiny \yng(2,1)}$. There are two possible ways of removing the
three boxes
$$\young({3}{2},{1})\qquad {\rm or}\qquad \young({3}{1},{2}).$$
Using the decomposition
$$ (123)=(12)(23),$$
the strand diagram gives
\begin{eqnarray} 
{1\over c_1-c_2}{1\over c_2-c_3} &=& -{1\over 2}\times 1 \quad {\rm for}\quad  \young({3}{2},{1})
\nonumber\\
&=&{1\over 2}\times -1\quad {\rm for}\quad \young({3}{1},{2}).
\label{firstresult}
\end{eqnarray}
The sum of these contributions is $-1$ as it should be. Refer to a particular order of removing the boxes from the
Young diagram as a {\sl path} through the Young diagram. The strand diagram computation for the character of the $m+n$ cycle
$\sigma_c$ in the hook representation $(n+m-s,1,1,\cdots,1)$ implies that
$$\sum_{\rm paths}\prod_{i=1}^{m+n-1}{1\over c_i-c_{i+1}}=(-1)^s .$$

First we establish that the restricted character of the $m+n$ cycle $\sigma_c$ with representation $(n+m-s,1,1,\cdots,1)$, and 
representation of the restriction obtained either by removing the box from the last or the first row. In this case, don't sum all 
paths - only sum paths that start from the last or first row respectively. Denote these two sums graphically as
$$\young({\,}{\,}{\,}{\,}{\,}{1},{\,},{\,},{\,})= 
\sum_{\rm paths\, starting\, in\, row\, 1}\prod_{i=1}^{m+n-1}{1\over c_i-c_{i+1}},$$
$$\young({\,}{\,}{\,}{\,}{\,}{\,},{\,},{\,},{1})= 
\sum_{\rm paths\, starting\, in\, row\, s+1}\prod_{i=1}^{m+n-1}{1\over c_i-c_{i+1}}.$$
The result we will establish says that (recall that there are $m+n$ boxes in the hook)
$$\young({\,}{\,}{\,}{\,}{\,}{1},{\,},{\,},{\,})= (-1)^s{m+n-1-s\over m+n-1},$$
$$\young({\,}{\,}{\,}{\,}{\,}{\,},{\,},{\,},{1})= (-1)^s{s\over m+n-1}.$$
Assuming this result is true for a hook with $m+n$ boxes and $s+1$ rows, it is straight forward to prove it is true for 
a hook with $m+n+1$ boxes and $s+1$ or $s+2$ rows. Indeed, for $m+n+1$ boxes in the hook and $m+n+1-s$ boxes in the first row
(so that the hook has $s+1$ rows), we have
\begin{eqnarray} 
\young({\,}{\,}{\,}{\,}{\,}{\,}{1},{\,},{\,},{\,})&=& 
\young({\,}{\,}{\,}{\,}{\,}{2}{1},{\,},{\,},{\,})+
\young({\,}{\,}{\,}{\,}{\,}{\,}{1},{\,},{\,},{2})\nonumber\\
&=&\young({\,}{\,}{\,}{\,}{\,}{1},{\,},{\,},{\,})+
{1\over m+n}\young({\,}{\,}{\,}{\,}{\,}{\,},{\,},{\,},{1})\nonumber\\
&=&{m+n-s-1\over m+n-1}(-1)^s +{1\over m+n}{s\over m+n-1}(-1)^s\nonumber\\
&=&{m+n-s\over m+n}(-1)^s\, .
\nonumber
\end{eqnarray}
For $m+n+1$ boxes in the hook and $m+n-s$ boxes in the first row (so that the hook has $s+2$ rows), we have
\begin{eqnarray} 
\young({\,}{\,}{\,}{\,}{\,}{\,},{\,},{\,},{\,},{1})&=& 
\young({\,}{\,}{\,}{\,}{\,}{2},{\,},{\,},{\,},{1})+
\young({\,}{\,}{\,}{\,}{\,}{\,},{\,},{\,},{2},{1})\nonumber\\
&=&-{1\over m+n}\young({\,}{\,}{\,}{\,}{\,}{1},{\,},{\,},{\,})
-\young({\,}{\,}{\,}{\,}{\,}{\,},{\,},{\,},{1})\nonumber\\
&=&-{1\over m+n}{m+n-s-1\over m+n-1}(-1)^s -{s\over m+n-1}(-1)^s\nonumber\\
&=&{s+1\over m+n}(-1)^{s+1}\, .
\nonumber
\end{eqnarray}
Thus, to establish the result it is enough to show that
$$ \young({\,}{1},{\,})=-{1\over 2}=\young({\,}{\,},{1}),$$
which has already been demonstrated in (\ref{firstresult}) above. 

We are now ready to tackle the computation of $\Tr_{(r_n,r_m)}\left(\Gamma_R (\sigma_c)\right)\, .$ We again use a strand diagram
technique, which amounts to summing over all paths and decomposing $\sigma_c$ into a product of two cycles. The values of all two 
cycles $(i,i+1)$ {\sl except for $(n,n+1)$} are given by $(c_i-c_{i+1})^{-1}$ with the weights $c_i$ read off the paths, coming 
from $r_n$ for $i\le n-1$ or from $r_m$ for $i>n$. If we strip off the boxes belonging to $r_n$ first, we can factor out a term which 
equals the character of $(1,2,\cdots,n)$ in irreducible representation $r_n$ so that $\Tr_{(r_n,r_m)}\left(\Gamma_R (\sigma_c)\right)$ 
vanishes unless $r_n$ is a hook. Stripping off the boxes that belong to $r_m$ first, allows us to conclude that 
$\Tr_{(r_n,r_m)}\left(\Gamma_R (\sigma_c)\right)$ again vanishes unless $r_m$ is a hook. Let the hook associated with $r_n$ have $n$ boxes,
arranged as $(n-s_n,1,1,\cdots,1)$ and let the hook associated with $r_m$ have $m$ boxes, arranged as $(m-s_m,1,1,\cdots,1)$. Given that 
both $r_n$ and $r_m$ have to be hooks, what are the allowed values of $R$? Further, given these values, what is 
$\Tr_{(r_n,r_m)}\left(\Gamma_R (\sigma_c)\right)\, $? The allowed values of $R$ fit into four possible cases. In what follows we will
list the structure of the $R,(r_n,r_m)$ label for these four cases.

{\vskip 0.5cm}

\noindent
{\it Case 1: $r_m$ contained in $r_n$ with no overlap:} To denote the structure of this case we display $R$ together with
an $x$ in the boxes which are removed to give $r_m$
$$\young({\,}{\,}{\,}{\,},{\,}{x}{x},{\,}{x},{\,}).$$
In general, the last box removed from $r_m$ does not correspond to a specific box in $R$. In this case the number $c_n-c_{n+1}$
appearing in the usual strand diagram computations is not well defined - the strand diagrams techniques of \cite{Bekker:2007ea}
are not applicable. However, for case 1, the box in the second row and second column is the last box of $r_m$ 
that is removed. Thus, $c_n-c_{n+1}$ has a definite value
and it is simple to compute the value of the $(n,n+1)$ cycle. Using the formulas given above, it is straight forward to verify that
\begin{eqnarray}
\Tr_{(r_n,r_m)}\left(\Gamma_R (\sigma_c)\right)&=&-(-1)^{s_m}{1\over s_n}{s_n(-1)^{s_n}\over n-1}
+(-1)^{s_m}{1\over n-s_n-1}{(n-s_n-1)(-1)^{s_n}\over n-1}\nonumber\\
&=&0.\nonumber
\end{eqnarray}

{\vskip 0.5cm}

\noindent
{\it Case 2: $r_m$ contained in $r_n$ with row overlap:} Display $R$ together with
an $x$ in the boxes which are removed to give $r_m$
$$\young({\,}{\,}{\,}{\,}{x},{\,}{x}{x},{\,}{x},{\,}).$$
In this case, the box in the second row and second column or the last marked box in the first row, is the last box of $r_m$ that is removed.
The argument for case 1 does not easily generalize to case 2 (or cases 3 and 4).

{\vskip 0.5cm}

\noindent
{\it Case 3: $r_m$ contained in $r_n$ with column overlap:} Display $R$ together with
an $x$ in the boxes which are removed to give $r_m$
$$\young({\,}{\,}{\,}{\,},{\,}{x}{x},{\,}{x},{\,},{x})$$

{\vskip 0.5cm}

\noindent
{\it Case 4: $r_m$ contained in $r_n$ with row and column overlap:} Display $R$ together with
an $x$ in the boxes which are removed to give $r_m$
$$\young({\,}{\,}{\,}{\,}{x},{\,}{x}{x},{\,}{x},{\,},{x})$$

{\vskip 0.5cm}

We will now give an argument applicable to all four cases. As a nontrivial check of our general formula, we will
verify that it predicts zero for case 1. Decompose our $n+m$ cycle $\sigma_c$ as
$$\sigma_c = (1,2,...,n)(n,n+1)(n+1,n+2,...,n+m).$$
Using the $S_n\times S_m$ symmetry enjoyed by our restricted character, we can replace
$$ (1,2,...,n)\to {\cal C}_n={\sum_{\rm n\,\, cycles}(i_1,i_2,...,i_n)\over (n-1)!},$$
$$ (n+1,n+2,...,n+m)\to {\cal C}_m={\sum_{\rm m\,\, cycles}(i_1,i_2,...,i_m)\over (m-1)!}.$$
${\cal C}_n$ is a sum over the $(n-1)!$ $n$-cycles in $S_n$; ${\cal C}_m$ is a sum over the $(m-1)!$ $m$-cycles in $S_m$.
It is clear that ${\cal C}_n$ and ${\cal C}_m$ are Casimirs of the $(r_n,r_m)$ representation.
Using the known characters of the hooks, it is straight forward to see that
$$ \Tr ({\cal C}_n )=(-1)^{s_n},\qquad \Tr ({\cal C}_m )=(-1)^{s_m},$$
so that these two Casimirs have eigenvalue
$$ {\cal C}_n={(-1)^{s_n}\over d_{r_n}},\qquad {\cal C}_m ={(-1)^{s_m}\over d_{r_m}}.$$
Thus, we now have
\begin{eqnarray} 
\Tr_{(r_n,r_m)}\left(\Gamma_R(\sigma_c)\right)&=&\Tr_{(r_n,r_m)}\left({\cal C}_n (n,n+1){\cal C}_m\right)\nonumber\\
&=&{(-1)^{s_n}\over d_{r_n}}{(-1)^{s_m}\over d_{r_m}}\Tr_{(r_n,r_m)}\left( (n,n+1)\right)\nonumber\\
&=&{(-1)^{s_n+s_m}\over mn}\left(\sum_i c_i -mN-{m(m-2s_m -1)\over 2}\right)\nonumber
\end{eqnarray}
where to get to the last line we have used formula (\ref{onblock}). The sum in the last line is over the weights in $R$ which are
peeled off and recombined to produce $r_m$. Notice that to evaluate these hooks we have not needed off the diagonal
block traces of the $(n,n+1)$ character, which is why we are able to obtain a simple and general formula. In addition, this formula
is in perfect agreement with the result obtained above for case 1.

\end{document}